\documentclass[a4paper,11pt]{article}
\usepackage{setspace}
\usepackage{feynmf}
\usepackage{amsmath,amssymb,amsthm}
\usepackage[totalwidth=17cm,totalheight=24cm]{geometry}
\usepackage{empheq}
\usepackage{graphicx}
\newcommand{\im}{{\mathrm i}}

\newcommand{\C}{{\mathbb C}}

\newcommand{\Hc}{{\cal H}}

\newcommand{\Tr}[1]{\mathrm{Tr}  \left( {#1}\right)}
\newcommand{\be}{\begin{eqnarray}}
\newcommand{\ee}{\end{eqnarray}}

\begin{document}

\title{Pure Connection Formalism for Gravity: Linearized Theory}
\author{Gianluca Delfino${}^1$, Kirill Krasnov${}^{1,2}$ and Carlos Scarinci${}^1$  \\ \\ \it{${}^1$ \, School of Mathematical Sciences, University of Nottingham}\\ \it{University Park, Nottingham, NG7 2RD, UK} \and 
\it{${}^2$ \, Max Planck Institute for Gravitational Physics}  \\ \it{Am M\"uhlenberg 1, 14476 Golm, Germany}}
\date{May 2012}
\maketitle
\begin{abstract}\noindent We give a description of gravitons in terms of an ${\rm SL}(2,\C)$ connection field. The gauge-theoretic Lagrangian for gravitons is simpler than the metric one, in particular because the Lagrangian only depends on 8 components of the field per spacetime point as compared to 10 in the Einstein-Hilbert case. Particular care is paid to the treatment of the reality conditions that guarantee that one is dealing with a system with a hermitian Hamiltonian. We give general arguments explaining why the connection cannot be taken to be real, and then describe a reality condition that relates the hermitian conjugate of the connection to its (second) derivative. This is quite analogous to the treatment of fermions where one describes them by a second-order in derivatives Klein-Gordon Lagrangian, with an additional first-order reality condition (Dirac equation) imposed. We find many other parallels with fermions, e.g. the fact that the action of parity on the connection is related to the hermitian conjugation. Our main result is the mode decomposition of the connection field, which is to be used in forthcoming works for computations of graviton scattering amplitudes. 
\end{abstract}

\section{Introduction}

Work \cite{Krasnov:2011pp} showed that $\Lambda\not=0$ General Relativity (GR) can be described in the "pure connection" formulation, in which the only dynamical field of the theory is a (complexified) ${\rm SO}(3)\sim{\rm SU}(2)$ connection rather than the metric.\footnote{Work \cite{Krasnov:2011pp} gave a gauge-theoretic description of a non-zero cosmological constant GR. Earlier works of Capovilla, Dell and Jacobson, see \cite{Capovilla:1989ac} and \cite{Capovilla:1991kx}, provide a similar description of the $\Lambda=0$ case. However, the action principle proposed in these works contains an additional auxiliary field on top of the connection. There is no need for such a field when $\Lambda\not= 0$, which results in literally a "pure connection" formulation.} Paper \cite{Krasnov:2011up} made the first steps towards setting up the perturbation theory in this formalism. In particular, the usual propagating degrees of freedom of GR (gravitons) were exhibited, and the propagator obtained. It was also shown that the same formalism is applicable to a very large class of (modified) gravity theories describing, as GR, just two propagating polarizations of the graviton. 

Here we develop this pure connection formalism for gravity further. This is the first in a series of papers aimed at studying how perturbative gravity can be described in this language. The principal aim of the present paper is to treat the linearized theory in the amount sufficient for later computations of e.g. graviton scattering amplitudes. However, interactions are considered only in the second paper of the series. 

In our treatment of the linearized theory particular attention is paid to the issues of the hermiticity of the arising quadratic Lagrangian. Indeed, as already mentioned, in the gauge-theoretic description of metric of Lorentzian signature one works with complexified ${\rm SU}(2)$, and thus ${\rm SL}(2,\C)$, connections. The Lagrangian then depends on the connection meromorphically, i.e. the complex conjugate of the connection field never enters. Such a description is only viable if some reality conditions are additionally imposed, and we discuss these in details in the present paper. Thus, our main results are the treatment of the hermiticity issues, as well as the related decomposition of the connection field into the modes. We also discuss the delicate issues of discrete C, P, T symmetries. The mode decomposition obtained in this paper gives everything that is needed for computations (performed in the second paper from the series) of graviton scattering amplitudes from the connection field correlation functions.  

Some aspects of our gauge-theoretic description of gravitons are quite unusual, and are therefore worth explaining already in the Introduction. To understand what is going on, it turns out to be particularly useful to use the language of (2-component) spinors. Before we explain how spinors appear in the pure connection description of gravity, let us remind the reader some very basic facts about them.

\subsection{Spinors}

We will necessarily be brief here, and send the reader to e.g. \cite{Penrose:1985jw} for more details. We recall that a tetrad $e$ is a map, at each spacetime point $p$, from the tangent space $T_p M$ to a copy of Minkowski space ${\cal M}^{1,3}$:
\be
e: T_pM\to {\cal M}^{1,3}.
\ee
The pull-back of the Minkowski metric $\eta$ on ${\cal M}^{1,3}$ gives the spacetime metric. Using the index notation we can write $g_{\mu\nu} = e_\mu^I  e_\nu^J \eta_{IJ}$, where $\mu,\ldots$ are the spacetime and $I,\ldots$ are "internal" indices, i.e. those referring to the Minkowski space ${\cal M}^{1,3}$ quantities. The object $\eta_{IJ}$ is the Minkowski metric, for which we choose the signature $(-,+,+,+)$. 

The spinors arise by introducing an identification between Minkowski vectors $x^I$ and $2\times 2$ (anti-) hermitian matrices
\be
{\bf x} := \im \left( \begin{array}{cc} x^0 + x^3 & x^1 -\im x^2 \\ x^1 + \im x^2 & x^0-x^3 \end{array} \right).
\ee
The Minkowski norm of $x^I$ is then expressed as the determinant of ${\bf x}$:
\be
-(x^0)^2+(x^1)^2+(x^2)^2+(x^3)^2 = {\rm det}({\bf x}).
\ee
It is then easy to see that the space of anti-hermitian matrices is preserved by the following action of the group ${\rm SL}(2,\C)$:
\be
{\bf x} \to g {\bf x} g^\dagger, \qquad g\in {\rm SL}(2,\C).
\ee
It is also clear that the above action preserves the determinant of $\bf x$ and thus the Minkowksi norm of the corresponding $x^I$. This provides an identification between the group ${\rm SL}(2,\C)$ and the Lorentz group ${\rm SO}(1,3)$:
\be
{\rm SO}(1,3)\sim {\rm SL}(2,\C).
\ee
The 2-component spinors are then objects that realize two inequivalent fundamental representations of the group ${\rm SL}(2,\C)$. Objects of one type, to which we shall refer as unprimed (using the GR terminology), transform simply as length 2 columns on which $g\in {\rm SL}(2,\C)$ acts by multiplication from the left. The objects of the second type (primed spinors) transform in a complex conjugate representation, and can be thought of as rows of length 2, on which $g^\dagger\in {\rm SL}(2,\C)$ acts from the right. Let us denote the space of spinors of unprimed type by $S_+$ and that of the opposite type spinors by $S_-$. Both spaces have an ${\rm SL}(2,\C)$-invariant "metric", which is however anti-symmetric, so that the norm of every object is zero. 

It is then clear that the matrix $\bf x$ is an object of a mixed type
\be
{\bf x} \in S_+ \otimes S_-.
\ee
Let us formalize this by introducing an index notation ${\bf x}^{AA'}$ for the matrix $\bf x$. Here $A,A'=1,2$ are the spinor indices, with an object of the type $\lambda^A\in S_+$ referred to as an unprimed spinor, and $\lambda^{A'}\in S_-$ as primed. Note that we can always identify the spinor spaces $S_\pm$ with their duals using the ${\rm SL}(2,\C)$-invariant metric. One must, however, be careful with the operation of raising and lowering of spinor indices, as this now introduces a minus sign (since the metric is anti-symmetric). We now write ${\bf x}^{AA'} := \im \sqrt{2} \, \theta_I^{AA'} x^I$, where we have introduced a matrix $\theta_I^{AA'}$ which is the object that fixes the identification between Minkowski vectors $x^I$ and anti-hermitian $2\times 2$ matrices $\bf x$. The factor of $\sqrt{2}$ is introduced for future convenience (so that the expression for $\theta_I^{AA'}$ in terms of the so-called doubly null tetrad is simple). The objects $\theta_I^{AA'}$ are hermitian: $(\theta_A^{AA'})^*=\theta_I^{AA'}$, where one also should take into account the fact that under the operation of complex conjugation the space of unprimed spinors goes into that of primed ones and vice-versa:
\be
(S_+)^*=S_-.
\ee
We can finally combine the tetrad $e_\mu^I$ with the object $\theta_I^{AA'}$ just introduced to form a new object $\theta_\mu^{AA'} = e_\mu^I \theta_I^{AA'}$ that is referred to as the {\it soldering form}. This object provides an identification between the space $S_+ \otimes S_-$ of mixed rank two spinors and the tangent space to our spacetime manifold $M$
\be
\theta: TM \to S_+\otimes S_-.
\ee
As $e$ that is used in its construction, it also carries information about the spacetime metric. The soldering form can be used to construct the {\it Dirac operator} $\nabla^{AA'} := \sqrt{2}\,\theta_\mu^{AA'} \nabla^\mu$, where $\nabla_\mu$ is the metric-compatible derivative operator, and we have raised the spacetime index on $\nabla$ using the metric. The Dirac operator, with its spinor indices raised or lowered appropriately using the ${\rm SL}(2,\C)$-invariant metrics on $S_\pm$ becomes a map sending spinors of one type into those of opposite type, e.g.:
\be
\nabla: S_+ \to S_-.
\ee
We are now ready to discuss the spinorial interpretation of the objects that appear in our gauge-theoretic formulation of gravity.

\subsection{${\rm SL}(2,\C)$ connections}

The main dynamical field of our theory is a complexified ${\rm SO}(3)\sim{\rm SU}(2)$ and thus ${\rm SL}(2,\C)$ connection. Locally its is a one-form on $M$ taking values in the Lie algebra ${\mathfrak g}\sim {\mathfrak sl}(2)$ of the gauge group. We will always think about the Lie algebra as a complex vector space of dimension 3. In index notations the connection is denoted by $A_\mu^i$, where $i=1,2,3$ is the Lie algebra index. As we shall see in details below, when the action of the theory is linearized around a suitable background connection, the background field allows for a certain metric to be defined. So, the linearized theory is about infinitesimal connections that we denote by $a_\mu^i$ living on a metric background. The metric allows us to define the usual notions of tetrad and then the spinors, as discussed above. We will then see that the structures available in the background field allow us to identify the Lie algebra ${\mathfrak g}$ with the space of symmetric rank 2 unprimed spinors
\be\label{g-s2}
{\mathfrak g} \sim S_+^2.
\ee
Indeed, as is well known, the Lie algebra ${\mathfrak sl}(2)$ of the Lorentz group (viewed as ${\rm SL}(2,\C)$), when considered as a complex vector space of dimension 3, is isomorphic to the second symmetric power of the fundamental representation. The background field then identifies the Lie algebra ${\mathfrak g}$ of the gauge group of the theory with the Lie algebra of the Lorentz group ${\mathfrak sl}(2)$ acting in each tangent space, and this is why (\ref{g-s2}) becomes possible.

Also, as we have already discussed, the spacetime index of our infinitesimal connection one-form can be converted into a pair of spinor indices using the soldering form $\theta_\mu^{AA'}$. Thus, overall, mapping all the indices of the infinitesimal connection into spinor ones we get an object
\be\label{intr-a}
a^{AA'\, BC}  \in S_+^2 \otimes S_+\otimes S_-.
\ee
Thus, our linearized theory is about fields living in the above spinor representation. This should be contrasted with the usual metric description where the metric perturbation field $h_{\mu\nu}$, when converted into the spinor form becomes
\be\label{intr-h}
h_{AA' BB'} \in (S_+\otimes S_-)\otimes_s (S_+\otimes S_-),
\ee
where $\otimes_s$ means the symmetric part of the tensor product. Both fields (\ref{intr-a}) and (\ref{intr-h}) are capable of describing a spin 2 particle (this follows just by counting the number of the fundamental spinor representations appearing, and multiplying the result by $1/2$, which is the spin carried by the fundamental representation). However, there is a profound difference between the two descriptions. The spinor space relevant for the usual metric description goes into itself under the operation of complex conjugation:
\be
( (S_+\otimes S_-)\otimes_s (S_+\otimes S_-))^*=(S_+\otimes S_-)\otimes_s (S_+\otimes S_-) .
\ee
However, the space in (\ref{intr-a}) under the operation of the complex conjugation gets sent to a completely different space
\be
(S_+^2 \otimes S_+\otimes S_-)^* = S_-^2 \otimes S_-\otimes S_+.
\ee
This is why there are real objects in the space in (\ref{intr-h}), but no real objects in the space in (\ref{intr-a}). In other words, the description of spin 2 particles is possible in terms of real fields if one uses fields such as $h_{\mu\nu}$, but cannot be possible if one uses the connection field in (\ref{intr-a}). This is the first conclusion that can be made about our prospective gauge-theoretic description of gravity even prior to developing it. As a result of this basic fact, the issues of reality conditions and hermiticity of the Lagrangian will have to be dealt with in a way significantly more non-trivial than in the metric based description, see more on this below. 

Let us ignore the issues of hermiticity for the moment, and discuss how the diffeomorphisms, which are the fundamental gauge symmetries of any theory of gravity, can be represented in our formalism. In the usual metric language the diffeomorphisms act via
\be\label{intr-diff}
\delta_\xi h_{\mu\nu} = \nabla_{(\mu} \xi_{\nu)},
\ee
where $\xi_\mu$ is the diffeomorphism generator. The important point about this transformation rule is that it involves the (first) derivatives of the generator. Therefore, the question of which components of $h_{\mu\nu}$ are pure gauge is mode-dependent, and can be answered only after the metric perturbation is decomposed into modes via an appropriate Fourier transform. The space (\ref{intr-h}) where metric perturbations live has dimension 10 (per spacetime point). The Hamiltonian analysis of gravity then tells us that 4 of the components of the metric perturbation field $h_{\mu\nu}$ get the interpretation of Lagrange multipliers imposing 4 constraints. This removes $4+4=8$ components, leaving only 2 propagating degrees of freedom of the graviton.

Let us now discuss a similar count of degrees of freedom in our gauge-theoretic description. The first fundamental difference is, as we shall see in details below, is that the connection transformation rule under the diffeomorphisms is much simpler than (\ref{intr-diff}). Thus, it turns out that the action of the diffeomorphisms is described by first decomposing the space in (\ref{intr-a}) into its two irreducible components
\be
S_+^2 \otimes S_+\otimes S_- = S_+^3 \otimes S_- \oplus S_+\otimes S_-,
\ee
where we have used the elementary representation theory fact that $S_+^2\otimes S_+= S_+^3\oplus S_+$. One then finds that from the two parts of the connection arising this way, the part taking values in $S_+\otimes S_-$ can be set to zero by an action of a diffeomorphism. In other words, $S_+\otimes S_-$ is pure gauge, and we can describe the space of (infinitesimal) connections ${\cal A}$ modulo diffeomorphisms in full generality as
\be\label{intr-a-diff}
{\cal A}/{\rm diffeos} = S_+^3\otimes S_-.
\ee
Importantly, this decomposition into a gauge and non-guage parts is mode-independent, and is possible already at the level of the Lagrangian, prior to any mode decomposition. This happens because it turns out to be possible to write the formula for the action of a diffeomorphism on the connection in a special way. Namely, in a gauge theory one has a freedom to talk about diffeomorphisms modulo the usual gauge transformations. Then one can write the formula for the infinitesimal diffeomorphism in such a way that it does not contain any derivatives of the generating vector field $\xi$. Explicitly, the action reads
\be\label{intr-diffeos}
\delta_\xi a^i_\mu = \xi^\nu F^i_{\mu\nu},
\ee
where $F^i_{\mu\nu}$ is the background curvature two-form. There are no derivatives of $\xi$ in this formula, and this is why the decomposition (\ref{intr-a-diff}) becomes possible. Below we shall see that the way that the decomposition (\ref{intr-a-diff}) is realized at the level of the action is that the Lagrangian is simply independent of the $S_+\otimes S_-$ components of the connection. 

To summarize, in our gauge-theoretic formulation, the diffeomorphisms are much easier to deal with than in the usual metric description. The components of the connection that are pure (diffeomorphism) gauge can be projected out already at the level of the Lagrangian and the action becomes a functional on the 8-dimensional space (\ref{intr-a-diff}). On this space one still has the usual ${\mathfrak sl}(2)$ gauge symmetries acting, with 3 of the 8 components of the projected connection field in (\ref{intr-a-diff}) being Lagrange multipliers for 3 constraints. At the end one gets the usual $8-3-3=2$ propagating modes of the graviton, but in a way completely different from the metric description. As we shall see below, in our description one will only need to gauge-fix the usual ${\mathfrak sl}(2)$ gauge symmetry, like one would be doing in Yang-Mills theory. In contrast, in the metric description one has to gauge-fix the diffeomorhisms, which leads to an arguably more involved formalism. Also to be emphasized, in our gauge theoretic description one will be dealing with only 8 components of the field per point, while in the metric description one has 10. Last but not least, as we shall see below, our gauge-fixed Lagrangian is actually a convex function in the field space, with all the modes having the same sign in front of their kinetic terms. This is not at all the case in the metric description, with one of the modes, namely the trace $h^\mu_\mu$, having an opposite sign in front of its kinetic term as compared to the other modes. This is the infamous conformal mode problem of the Euclidean approach to quantum gravity. This problem is absent in the present gauge-theoretic formulation of gravity, with the Euclidean signature Lagrangian (when all the fields become real) being a non-negative (i.e. convex in non-flat directions) function in the field space. This fact, as well as other simplifications resulting from the possibility to project away the diffeomorphisms from the outset, should be viewed as the main reason for taking the present gauge-theoretic formulation as a serious alternative to the usual metric-based one. We refer the reader to \cite{Krasnov:2012pd} for a further discussion of the above points.

\subsection{Fermions}

Above we have seen that our infinitesimal connection field cannot be real, as it takes values in a space that does not go into itself under the complex conjugation. Of course, the full complex-valued field then describes twice more real modes than is needed (with the extra half of the modes coming from the complexification badly behaving). Thus, one does need to impose some reality conditions if one wants to get a satisfactory description of spin 2 particles. The way this happens turns out to be strongly analogous to what happens in theories of fermions, i.e. spin 1/2 particles. Thus, let us briefly discuss the usual fermions in Minkowski spacetime first.

A possible (and in fact rather powerful, but not commonly known) approach to fermions is to describe them by a second-order in derivatives action, treating the original Dirac first-order equation as a reality condition for the fermion field. This gives a completely equivalent description to the usual one, and can also be shown to lead to some simplifications in the computations of Feynman diagrams, see e.g. \cite{Chalmers:1997ui} for an emphasis of this fact. 

To describe this in some details, let us only discuss here the case of a single Majorana fermion, which is the simplest (and is also enough for our purposes of drawing an analogy). In the usual first-order Dirac like formulation this is described by the Lagrangian
\be\label{Maj-L}
{\cal L}_{\text Majorana} = \im \sqrt{2} \lambda^\dagger_{A'} \theta_\mu^{AA'} \partial^\mu \lambda_{A} - (m/2) \lambda^A \lambda_A - (m/2) \lambda^\dagger_{A'} \lambda^{\dagger A'},
\ee
where $\lambda_A,\lambda^\dagger_{A'}$ are two anti-commuting 2-component spinors and $\lambda^\dagger_{A'}$ is the hermitian conjugate of $\lambda_A$. The above Lagrangian is hermitian modulo a surface term, as can be checked by an easy computation. 

In the second-order description one integrates out the primed spinors $\lambda^\dagger_{A'}$ (using the fact that at the level of the path integral it is legitimate to treat $\lambda_A,\lambda^\dagger_{A'}$ as independent fields. To do this one uses the field equation for $\lambda^\dagger_{A'}$ that reads:
\be\label{reality}
\lambda^{\dagger A'} = \frac{\im \sqrt{2}}{m} \theta_\mu^{AA'} \partial^\mu \lambda_A.
\ee
One then substitutes this back into (\ref{Maj-L}) to obtain (after using some algebra of soldering forms)
\be\label{Maj-2-order}
{\cal L}_{\text Majorana} = -\frac{1}{2m} \partial^\mu \lambda^A \partial_\mu \lambda_A - \frac{m}{2} \lambda^A \lambda_A,
\ee
which is just the Lagrangian that gives the Klein-Gordon equation for each of the two components of $\lambda_A$. It can then be shown that the theory (\ref{Maj-2-order}) supplemented with the reality conditions (\ref{reality}) is completely equivalent to the original theory (\ref{Maj-L}). Of course, the Lagrangian (\ref{Maj-2-order}) is not hermitian, but instead depends holomorphically on the spinor field $\lambda_A$. It only leads to a theory with a hermitian Hamiltonian once the theory is restricted to live on the space of fields satisfying (\ref{reality}). There are some subtle points here about on-shell versus off-shell correspondence, and this will be further discussed in the main text, when contrasting with what happens in our gauge-theoretic description. 

It is worth discussing the reality condition (\ref{reality}) from a more general viewpoint. Imagine we would like to start with (\ref{Maj-2-order}), and then find some appropriate reality condition that would give us a theory with a hermitian Hamiltonian. The spinor field $\lambda_A$ that we work with lives in the space $S_+$, and this space goes into $S_-$ under the complex conjugation. Thus, the field cannot be taken to be real. We then need a more sophisticated real structure on the complex phase space of our theory, and this is provided by the Dirac operator. Indeed, the Dirac operator maps spinors of one type into those of the other. Thus, we can combine the action of the Dirac operator with that of the complex (hermitian) conjugation to define
\be\label{R}
{\cal R} : = \frac{1}{\im\, m} \partial \circ \dagger,
\ee
where $\partial$ here stands schematically for the Dirac operator as we defined it above. The $\cal R$-operator is an anti-linear map sending the space of unprimed spinors into itself. Importantly, it becomes an involution ${\cal R}^2={\rm Id}$ on the space of solutions of the theory (\ref{Maj-2-order}), and is thus a real structure on the phase space when the latter is viewed as the space of solutions of field equations. The reality condition (\ref{reality}) is then just the condition selecting the real section of the phase space with respect to the real structure $\cal R$. This gives an equivalent viewpoint on the usual theory of fermions that works with first-order hermitian Lagrangians, but also leads to some important simplifications in computations with fermions, as is emphasized in \cite{Chalmers:1997ui}. So, this is a valid viewpoint on the fermions. As we now discuss, gravitons in their gauge-theoretic formulation share many similarities with this description of fermions. 

\subsection{Reality for gauge-theoretic gravitons}

We now come back to the description of the gravitons as connections taking values in (\ref{intr-a}), or, after the diffeomorphism components have been projected away, in (\ref{intr-a-diff}). As we shall see, the resulting linearized Lagrangian on this space is a meromorphic function of the connection, leading to a second-order in derivatives field equation. Since the connection takes values in $S_+^3\otimes S_-$, and this space is not invariant under the operation of complex conjugation, the connection cannot be real. However, we can now use the above second-order treatment of fermions as a guide, and device an appropriate reality condition that will make the Hamiltonian hermitian. 

The idea is to cook up an anti-linear map from the space $S_+^3\otimes S_-$ into itself by combining the operation of the hermitian conjugation of the field with the action of an appropriate differential operator. The operator that we have at our disposal is the Dirac operator $\nabla$. Note that we now work in a curved background, and so refer to the Dirac operator as $\nabla$ in contrast to $\partial$ above. The importance of the curved background will be explained below. The Dirac operator converts one spinor index into the index of an opposite type. Thus, if we take the complex conjugate of an object in $S_+^3\otimes S_-$ we get an object in $S_-^3\otimes S_+$. To convert this into an object in the original space $S_+^3\otimes S_-$ we need to flip two of the spaces $S_-$ to become $S_+$. Thus, we will have to apply the Dirac operator twice. In other words, a possible reality condition must be of the form
\be\label{R-2}
{\cal R} \sim \nabla^2\circ \dagger.
\ee
We now note that in the case of the Dirac theory we had the mass parameter that allowed to make the dimensions match in (\ref{R}), so that $\cal R$ is a dimensionless operator. For the graviton there is clearly no mass parameter that can be used, as the graviton is massless. It is for this reason that our description of gravitons only makes sense in a curved background, where the radius of curvature of the background can provide the missing dimensionful parameter. This provides yet another explanation of why the gauge-thereotic description of gravity only works properly when $\Lambda\not=0$. Below we shall see that it is the mass parameter associated with the curvature $M^2\sim \Lambda$ whose inverse power will be sitting in (\ref{R-2}) to make the dimensions match. We will also see that, on solutions of field equations, an appropriately designed anti-linear operator of the form (\ref{R-2}) becomes an involution, and thus defines a real structure on the space of solutions (=phase space). After the corresponding real section is selected, one obtains a theory with a hermitian Hamiltonian. In fact, as we shall also demonstrate, the corresponding complex description of the phase space of gravitons is just a (complex) canonical transformation of the usual phase space in terms of the metric perturbation. So, at the level of the (reduced) phase space the two descriptions will be shown to be completely equivalent. 

We summarize by saying that our gauge-theretic description (to be developed in the main text) is completely equivalent to the standard description at the level of the fully symmetry reduced phase space. However, the connection viewpoint on gravitons brings some important simplifications into the perturbation theory, as could be suspected from the fact that the theory now depends on less components of the field to start from (8 as compared to 10). A related fact is that in the gauge-theoretic description the field takes values (after the diffeomorphisms have been dealt with as in (\ref{intr-a-diff})) in an irreducible representation $S_+^3\otimes S_-$ of the Lorentz group. This is in contrast to the usual description, where one must build up the perturbation theory working with all the components of the metric perturbation. These split into two irreducible components $S_+^2\otimes S^2_-$ and the trivial representation (functions on spacetime). The two irreducible components behave very differently, and part of the complexity of the standard perturbation theory consists in dealing with these two different components. This problem is absent in our treatment, and will be seen to result in many simplifications in the formalism.

\bigskip

Now that we have explained the main unusual points of our construction, we can start with our development of the diffeomorphism invariant ${\rm SO}(3)\sim {\rm SU}(2)$ gauge theory, which will be shown to describe gravity. We start with a formulation of the theory in Section \ref{sec:theory}. We then discuss the background and obtain the linearized Lagrangian in Section \ref{sec:lin}. The resulting free theory is described in details in Section \ref{sec:free}, where also the Hamiltonian analysis is performed. Section \ref{sec:reality} is central for the whole story and discusses the subtle points related to the reality conditions. It also introduces the metric variable, in terms of which one has the familiar dynamics. Section \ref{sec:metric} shows that the passage to the metric variable is a canonical transformation on the phase space of the theory. The mode decomposition is obtained in Section \ref{sec:modes}, and then the discrete symmetries are discussed in Section \ref{sec:discrete}. We conclude with a discussion.

\section{The theory}
\label{sec:theory}

The contents of this section are not new. Some more details on diffeomorphism invariant gauge theories described below can be found in \cite{Krasnov:2012pd}. General Relativity (with $\Lambda\not=0$) was first formulated in this language in \cite{Krasnov:2011pp}. 

\subsection{Diffeomorphism invariant gauge theories}

We begin in full generality, and define a large class of what can be called diffeomorphism invariant gauge theories for an arbitrary gauge group. Thus, let $G$ be a (complex) Lie group, which we for simplicity assume here to be simple. Consider a $G$-connection on the spacetime $M$. Locally it can be described as a one-form $A_\mu^I$ with values in the Lie algebra $\mathfrak g$ of $G$. Thus, here and in what follows $I=1,\ldots, n$ is the Lie algebra index. The curvature of the connection is a two-forms with values in $\mathfrak g$ that can be described as
\be
F^I = dA^I + \frac{1}{2}f^I{}_{JK} A^J\wedge A^K,
\ee
where $f^I{}_{JK}$ are the structure constants. 

Now let $f$ be a scalar valued function acting on symmetric matrices in ${\mathfrak g}\otimes_s{\mathfrak g}$:
\be
f: {\mathfrak g}\otimes_s{\mathfrak g}\to \C.
\ee
We require this function to satisfy two properties: (i) It must be gauge invariant $f({\rm Ad}_g X)=f(X), \forall g\in G$; (ii) It must be homogeneous of degree one $f(\alpha X)=\alpha f(X), \forall\alpha\not=0$. Both conditions are required to hold for any $X\in{\mathfrak g}\otimes_s{\mathfrak g}$. 

Having such a function, it is not hard to see that it can be applied to the quantity $F^I\wedge F^J$, with the result being a well-defined 4-form. Indeed, $F^I\wedge F^J\in \Lambda^4 \otimes {\mathfrak g}\otimes_s{\mathfrak g}$, i.e. it is a 4-form with values in the space of symmetric matrices. We can apply the function $f$ to it, and the result is gauge-invariant due to the gauge-invariance of $f$. At the same time, the 4-form factor can be just "taken out" from the function due to its homogeneity, and so one gets a well-defined 4-form. Integrating this over the manifold one gets the action
\be\label{action}
S[A]= \im \int_M f(F\wedge F).
\ee
Several remarks about this action are in order. First, the factor of $\im=\sqrt{-1}$ is introduced for future convenience. Second, there are no dimensionful coupling constants in our theory. Indeed, there are only dimensionless parameters involved in constructing the function $f$. All the dimensions are carried by the fields, so that the connection $A$ has the mass dimension one, and the curvature has the mass dimension 2. The Lagrangian then has the required mass dimension 4 by the homogeneity of $f$. Below we shall see that the dimensionful coupling constants get introduced into this theory when a suitable background is selected (as combinations of the mass scale of the background with the other dimensionless parameters present in $f$). 

Another remarks about (\ref{action}) is that its field equations are of the second order in derivatives. This is easy to see if we write the equations as
\be\label{feqs-general}
d_A B^I = 0, \qquad {\rm where} \qquad B^I:= \frac{\partial f}{\partial X^{IJ}} F^J,
\ee
and where the matrix $X^{IJ}=F^I\wedge F^J$. As we shall see below, the matrix of derivatives of $f$ with respect to $X^{IJ}$ is a well-defined matrix-valued function (not a form) of homogeneity degree zero acting on $\Lambda^4\otimes {\mathfrak g}\otimes_s{\mathfrak g}$. Thus, the quantity $B^I$ is a well-defined 2-form with values in the Lie algebra. The field equations are then just a statement that $B^I$ is covariantly constant with respect to the connection $A$. Let us now count the number of the derivatives appearing in the equation (\ref{feqs-general}). The function $f$, as well as the matrix of its first derivatives, are (highly non-linear) functions of the first derivatives of $A$. Then another derivative is taken in (\ref{feqs-general}), which results in second-order field equations. 

Our last remark about (\ref{action}) is that for a generic $f$ they are dynamically non-trivial theories, i.e. describe propagating degrees of freedom. The clause about generic $f$ is important, for there is one point in the theory space corresponding to $f(F\wedge F)={\rm Tr}(F\wedge F)$ which gives a topological theory without any propagating modes. But this is clearly a very special point in the theory space because, as we shall see below, whenever the Hessian of the function $f$ is non-degenerate there are propagating modes. For a generic $f$ it can be shown by a Hamiltonian analysis, see \cite{TorresGomez:2009gs} for such an analysis in a different, but related description, that the theory (\ref{action}) describes $2n-4$ propagating modes. 

\subsection{Gravity}

It turns out \cite{Krasnov:2011up} (and this will be shown below) that when one takes $G={\rm SL}(2,\C)$, viewed as a 3-dimensional complex Lie group (i.e. as a complexification of ${\rm SU}(2)$), the above theory describes, for any choice of the defining function $f$, interacting massless spin 2 particles. This statement does not take into account the reality conditions issues, as discussed in the Introduction. In other words, we do not know if there is a choice of the reality conditions that render a theory with arbitrary $f$ to have a hermitian Hamiltonian. However, what we will show in this paper is that, when linearized around an appropriate background (which is going to be just de Sitter space in the language of connections), all theories (\ref{action}) with $G={\rm SL}(2,\C)$ lead to the same linearized dynamics. This dynamics is that of massless spin 2 particles, and then (linearized) reality conditions can be imposed to yield a positive-definite hermitian Hamiltonian. Thus, there is a satisfactory treatment of the reality conditions issue at the linearized level for any $f$. Whether this can be extended to the full non-linear level is an open problem, apart from the case of $f$ that corresponds to GR, where the correspondence to GR implies that there is a satisfactory solution to the reality conditions problem. 

\subsection{General Relativity}

General Relativity with a non-zero cosmological constant can also be described in this language, and is just a particular point in the theory space (\ref{action}). In this case the action reads, see \cite{Krasnov:2011pp}
\be\label{GR}
S_{\rm GR}[A]=\frac{\im}{16\pi G \Lambda}  \int \left( \mathrm{Tr}{ \sqrt{F\wedge F} } \right)^2,
\ee
where $G$ is the usual Newton's constant, $\Lambda$ is the cosmological constant, $\im=\sqrt{-1}$, and $F^i=d A^i +(1/2)\epsilon^{ijk} A^j\wedge A^k$ is the curvature of $A^i$. Due to the presence of the factors of imaginary unit in front of the action, and also because of the fact that the connection is complex (reality conditions will be describe below), it is not obvious that this action describes a theory with unitary dynamics. Still, as we shall see in particular from the graviton scattering results (in the second paper from the series), it describes the usual general relativity. An argument establishing equivalence to the usual metric based GR at the full non-linear level is given in \cite{Krasnov:2011pp}. Thus, we know for sure that at least for one of the members of the class (\ref{action}) the issue of reality conditions at the full non-linear level can be dealt with satisfactorily (by going to the usual metric-based real description). 

The square root of a matrix appearing in the action has to be understood perturbatively, as we shall explain (and explore) below. Note that the Newton's constant appears in front of the action only in the dimensionless combination $G\Lambda$. This is of course also possible in the usual metric-based formulation if one rescales the metric to absorb $\Lambda$ into the volume factor $\sqrt{-g}$. The metric then becomes dimensionful and $\Lambda$ appears in front of the action exactly as in (\ref{GR}). Our final remark about (\ref{GR}) is that it gives only an on-shell equivalent formulation of general relativity, while off-shell the action (\ref{GR}) has different convexity properties from the Einstein-Hilbert one. This is of no importance for the present  and the second paper from the series, where only the tree-level scatting amplitudes are studied, since these can be expected to be the same as in GR. However, one should be cautious when comparing the (to be constructed) quantum theory based on (\ref{GR}) with the one based on the Einstein-Hilbert functional. Even though the phase spaces of both theories are the same (viewed as the spaces of solutions of field equations), there can be subtleties (e.g. in the measure) when comparing the path-integral based quantum theories. We do not touch these issues any further in the present work.

\section{Perturbative expansion}
\label{sec:lin}

The treatment of the background below is along the lines of \cite{Krasnov:2011hi}. A more in depth discussion of the mass scale introduced by the background in available in \cite{Krasnov:2012pd}. The perturbative expansion of the action is to a large extent new, with only a very preliminary discussion available in \cite{Krasnov:2011up}.

\subsection{The background}

We are (eventually) interested in developing Feynman rules for the theories (\ref{action}), and, in particular, for (\ref{GR}). One immediate difference with the case of the metric-based GR is that we cannot directly expand around a background that corresponds to the Minkowski spacetime. Indeed, our action (\ref{GR}), strictly speaking, only describes the $\Lambda\not=0$ situation, as it blows up if one sends $\Lambda\to 0$. Thus, the best we can do (if we are after the Minkowski spacetime scattering amplitudes) is to expand around a constant curvature background and take the curvature scalar to zero at the end of the calculation. This is the strategy that will be followed here (and was previously followed in \cite{Krasnov:2011up}). As we shall see below, the presence of the cosmological constant at intermediate stages of the computations will make to us available constructions that are simply impossible in the usual metric setting of zero $\Lambda$. 

We shall consider perturbations around a fixed constant curvature background connection. To explain what constant curvature means in our setting let us start by describing a general homogeneous and isotropic in space ${\rm SO}(3)$ connection. First, a general homogeneous in space connection is of the form
\be
A^i= a^{ij}(\eta) dx^j + b^i(\eta) d\eta,
\ee
where we have indicated that the components can only be functions of the time coordinate $\eta$. It is obvious that we can kill the $b^i(\eta)$ components by a time-dependent gauge transformation. This leaves us with the first term only. We now require that the effect of an ${\rm SO}(3)$ rotation of the coordinates $x^i$ (around an arbitrary center) can be offset by an ${\rm SO}(3)$ gauge transformation. This implies that $a^{ij}$ must be proportional to $\delta^{ij}$ for all $\eta$. Thus, we are led to consider the following connections:
\be\label{A*}
A^i = \frac{c(\eta)}{\im} dx^i,
\ee
where the function $c(\eta)$ is arbitrary, and we have introduced a factor for $1/\im$ for future convenience. We now note that the curvature of this connection is given by
\be\label{A-curv*}
F^i = \frac{c'}{\im} d\eta\wedge dx^i - \frac{c^2}{2} \epsilon^{ijk} dx^j \wedge dx^k,
\ee
where the prime denotes the derivative with respect to $\eta$. This means that we have
\be
F^i\wedge F^j\sim \delta^{ij}.
\ee
Thus, for our chosen background (\ref{A*}) the matrix $\tilde{X}^{ij}$ is proportional to the identity matrix, which means that the matrix of first derivatives of the function $f(\tilde{X})$ is also proportional to the identity. This implies that any connection (\ref{A*}) satisfies the field equations following from (\ref{action})
\be
D_A\left( \frac{\partial f}{\partial \tilde{X}^{ij}} F^j \right) =0,
\ee
as these equations reduce to the Bianchi identity $D_A F^i=0$. This happens for any $f$, i.e. for any of the theories in our theory space. 

We now note that the curvature (\ref{A-curv*}) can be written as
\be
F^i = - c^2 \left(\frac{\im c'}{c^2} d\eta\wedge dx^i + \frac{1}{2} \epsilon^{ijk} dx^j \wedge dx^k\right).
\ee
We can now choose the time coordinate conveniently, so that
\be\label{eta-t}
\frac{c'}{c^2} d\eta = dt,
\ee
and then write
\be
F^i = - c^2 \left( \im dt\wedge dx^i + \frac{1}{2} \epsilon^{ijk} dx^j\wedge dx^k\right),
\ee
where $c$ should now be thought of as a function of $t$. In fact, from (\ref{eta-t}) we have $dc/dt=c^2$ and thus
\be
c(t)=-\frac{1}{t-t_0},
\ee
where $t_0$ is the integration constant. All in all, we see that, by an appropriate choice of the $t$ coordinate, we can rewrite the curvature of any of the connections (\ref{A*}) as
\be\label{F-S}
F^i=-M^2 \Sigma^i,
\ee
where 
\be\label{Sigmas}
\Sigma^i = a^2 \left( \im dt\wedge dx^i + \frac{1}{2} \epsilon^{ijk} dx^j\wedge dx^k\right)
\ee
are the self-dual two-forms for the de Sitter metric
\be\label{dS-metric}
ds^2 = a^2 \left( -dt^2 + \sum_i (dx^i)^2\right),
\ee
and
\be
a(t) = - \frac{1}{M(t-t_0)}
\ee
is the usual de Sitter scale factor as a function of the (conformal) time $t$. Note that we have introduced an arbitrary dimensionful parameter $M$ in (\ref{F-S}). This parameter is directly related to the radius of curvature of the de Sitter metric (\ref{dS-metric}). It is completely arbitrary, as we can always rescale both $M$ and $\Sigma^i$ in (\ref{F-S}) without changing the curvature. But once introduced, it determines the metric, and thus determines how all scales in the theory are measured. The condition (\ref{F-S}), which as we saw can be always achieved by choosing the time coordinate appropriately, is our constant curvature condition for the background connection. The essence of this condition is that it introduces a (background) metric into our background-free up to now description, and fixes how all scales are measured. 

It is worth discussing the construction that introduced a metric into our so far metric-free story in more details. This is a geometrical construction known for many years, and is in particular due to \cite{Urbantke:1984eb}. The idea is that when the triple of curvatures $F^i$ of the connection $A^i$ is linearly independent, the 3-dimensional space that it spans in the space of all 2-forms can be declared to be the space of self-dual 2-forms for some metric. It is then known that this determines the metric modulo conformal transformations. This is precisely how the metric (\ref{dS-metric}) appeared from the background connection (\ref{A*}). We have also made a further choice of the conformal factor by so that the connection becomes one of constant curvature in the sense of equation (\ref{F-S}). Fixing $M$ in that equation to be constant eliminates the conformal freedom in the choice of the metric, up to constant rescalings. A choice of a particular constant $M^2$ in that equation is then equivalent to a choice of units in which all other quantities in our theory are measured. In this sense $M$ is not a parameter of the theory, it is rather a scale in terms of which all other scales in the theory get expressed. Thus, e.g. in the second paper in the series we shall see how the gravitons' interaction strength (Newton constant) appears as constructed out of $M$ and the dimensionless coupling constants present in our theory. 

\subsection{Working with functions of matrix-valued 4-forms}

We should now explain how a function (e.g. the square root in (\ref{GR})) can be applied to forms. We do this in a way most convenient for practical compuations. Thus, it is convenient to use a completely anti-symmetric density $\tilde{\epsilon}^{\mu\nu\rho\sigma}$ available without any metric to construct the following densitiezed matrix:
\be\label{X}
\tilde{X}^{ij}=\frac{1}{4} \tilde{\epsilon}^{\mu\nu\rho\sigma} F^i_{\mu\nu}F^j_{\rho\sigma}.
\ee
The general action (\ref{action}) for $G={\rm SL}(2,\C)$ then bomes
\be\label{action-general}
S[A] = \im \int d^4x\, f(\tilde{X}^{ij}).
\ee
One can now see that the integrand is a density weight one scalar, and so the integral is well-defined. The field equations then take the form
\be
d_A \left( \frac{\partial f}{\partial \tilde{X}^{ij}} F^j\right) = 0,
\ee
where the matrix of first derivatives that appears is now just that of usual derivatives of a function of a matrix with respect to the matrix components. For GR action (\ref{GR}) written in terms of $\tilde{X}^{ij}$ we have:
\be\label{GR'}
S_{\rm GR}[A] = \frac{\im M_p^2}{3M^2} \int d^4 x \left( \mathrm{Tr}{ \sqrt{\tilde{X}} } \right)^2.
\ee
Here we have introduced $M_p^2:=1/16\pi G, M^2:=\Lambda/3$. What we have now is the square root of a symmetric $3\times 3$ matrix, and this is well-defined (at least for matrices that are not too far from the identity matrix). The action in the form (\ref{GR'}) will be our starting point for developing the GR perturbation theory (in the second paper from the series). 

\subsection{A convenient way to write the action}

Let us now consider the value of $\tilde{X}^{ij}$ at the background. We have
\be
\tilde{X}^{ij}\,\hat{=}\, \frac{M^4}{4} \ \tilde{\epsilon}^{\mu\nu\rho\sigma}\  \Sigma^i_{\mu\nu} \Sigma^j_{\rho\sigma} = 2\im M^4 \sqrt{-g}\,\, \delta^{ij},
\ee
where our convention is that the hat means "evaluated at the background". Here we made use of the self-duality of $\Sigma$'s and the algebra (\ref{S-algebra}) of $\Sigma$'s. It is very convenient to rescale the $\tilde X$ variable by $2\im M^4 \sqrt{-g}$ so that the result equals to the Kronecker delta on the background. Thus, we introduce:
\be
\hat{X}^{ij}:=\frac{\tilde{X}^{ij}}{2\im M^4 \sqrt{-g}} \,\hat{=}\, \delta^{ij}.
\ee
We now rewrite the general gravity action (\ref{action-general}) in terms of $\hat{X}$. We have:
\be\label{action-general'}
S[A] = -2M^4 \int d^4x \sqrt{-g}\, f(\hat{X}^{ij}).
\ee
For the GR action (\ref{GR'}) this becomes:
\be\label{GR''}
S_{\rm GR}[A]=-\frac{2}{3} M^2_p M^2  \int d^4x \sqrt{-g}\, \left( \mathrm{Tr}{\sqrt{\hat{X}^{ij}}} \right)^2.
\ee
It then becomes a simple exercise to compute the variations of the action, see below. 

\subsection{Evaluating action at the background}

Let also discuss the value of the actions (\ref{action-general'}) and (\ref{GR''}) when evaluated on the background. We have, for the general action:
\be\label{action-backgr-general}
S[A]\hat{=} -2M^4 f(\delta) \int d^4x \sqrt{-g}\,.
\ee
For (\ref{GR''}) this becomes
\be
S_{\rm GR}[A]\hat{=} - 6 M_p^2 M^2 \int d^4x \sqrt{-g}\, = -\frac{\Lambda}{8\pi G}\int d^4x \sqrt{-g}\,,
\ee
which is the same as the value of the Einstein-Hilbert action 
\be
S_{\rm EH}[g] = - \frac{1}{16\pi G}\int d^4x \sqrt{-g}\, (R-2\Lambda)
\ee
evaluated on the de Sitter metric (\ref{dS-metric}). We see from (\ref{action-backgr-general}) that for a general theory the dimensionless quantity $f(\delta)$ plays the role of a combination $3M_p^2/M^2$ in the case of GR. We emphasize, however, that for a general theory there is no notion of the Planck constant, at least not until graviton interactions are considered. In the second paper of the series we compute the graviton interactions strength and will extract an appropriate dimensionful coupling constant this way. It is however, not guaranteed that the Planck mass obtained from this Newton constant will be related with the dimensionless parameter $f(\delta)$ in front of the background-evaluated action in exactly the same way as in GR. 

\subsection{Variations}

We start by computing the variations of $\hat X$, as a function of the connection, evaluated at the background $\hat{X}^{ij}\,\hat{=}\,\delta^{ij}$. We have:
\be\label{X_variations}
\delta\hat{X}^{ij}\,\hat{=}\,-\frac{1}{M^2}\Sigma^{(i\mu\nu}D_\mu\delta A^{j)}_\nu,
\\\nonumber
\delta^2 \hat{X}^{ij}\,\hat{=}\,\frac{1}{\im M^4}\epsilon^{\mu\nu\rho\sigma}D_\mu\delta A^i_{\nu}D_\rho\delta A^j_\sigma-\frac{1}{M^2}\Sigma^{(i\mu\nu} \epsilon^{j)kl}\delta A_\mu^k   \delta A_\nu^l,
\\\nonumber
\delta^3 \hat{X}^{ij}\,\hat{=}\,\frac{3}{\im M^4}\epsilon^{\mu\nu\rho\sigma}D_\mu\delta A_\nu^{(i}\epsilon^{j)kl}\delta A_\rho^k\delta A^l_\sigma.
\ee
Finally, the fourth variation is zero $\delta^4 \hat{X}^{ij}=0$ even away from the background. In all expressions above $D_\mu$ is the covariant derivative with respect to the background connection. Thus, it is important to keep in mind that $D$'s do not commute:
\be\label{comm-DD}
2 D_{[\mu} D_{\nu]} V^i = \epsilon^{ijk} F_{\mu\nu}^j V^k,
\ee
for an arbitrary Lie algebra valued function $V^i$. Here $F_{\mu\nu}^i$ is the background curvature (\ref{F-S}). Thus, the commutator (\ref{comm-DD}) is of the order $M^2$. This has to be kept in mind when (in the limit $M\to 0$) replacing the covariant derivatives $D$ with the usual partial derivatives.

\subsection{Variations of the general action}

We will now explain a procedure that can be used for computing the perturbative expansion of the action (\ref{action-general'}). It is completely algorithmic, and is not hard to implement to an arbitrary order. In this paper we will only need the second variation, but we decided to explain the general procedure already here since once the general principle is understood, it is not hard to implement to get the interactions as well. First, les us define a convenient notation 
$$f^{(n)}_{ijkl...}=\left.\frac{\partial^nf}{\partial\hat X^{ij}\partial\hat X^{kl}...}\right|_{\delta},$$
where the derivatives are all evaluated at the background $\hat{X}^{ij}=\delta^{ij}$. The variations of the action are then given by:
\be\label{variations-action-general}
\delta S\,\hat{=}\,-2M^4\int f^{(1)}_{ij}\delta\hat X^{ij},\qquad 
\delta^2 S\,\hat{=}\,-2M^4\int \left[f^{(2)}_{ijkl}\delta\hat X^{ij}\delta\hat X^{kl}+f^{(1)}_{ij}\delta^2\hat X^{ij}\right],
\\ \nonumber
\delta^3 S\,\hat{=}\,-2M^4\int \left[f^{(3)}_{ijklmn}\delta\hat X^{ij}\delta\hat X^{kl}\delta\hat X^{mn}+3f^{(2)}_{ijkl}\delta^2\hat X^{ij}\delta\hat X^{kl}+f^{(1)}_{ij}\delta^3\hat X^{ij}\right],
\\ \nonumber
\delta^4 S\,\hat{=}\,-2M^4\int \left[f^{(4)}_{ijklmnpq}\delta\hat X^{ij}\delta\hat X^{kl}\delta\hat X^{mn}\delta\hat X^{pq}+6f^{(3)}_{ijklmn}\delta^2\hat X^{ij}\delta\hat X^{kl}\delta\hat X^{mn}\right.
\\ \nonumber
\left.+4f^{(2)}_{ijkl}\delta^3\hat X^{ij}\delta\hat X^{kl}+3f^{(2)}_{ijkl}\delta^2\hat X^{ij}\delta^2\hat X^{kl}\right].
\ee
Below we shall explain how the derivative matrices appearing here can be parameterized conveniently. However, let us first consider the special case of the GR action. 

\subsection{Variations of the GR action}

For the case of GR we have
\be
f_{\rm GR}(\hat X)=\frac{M_p^2}{3M^2} \Tr{\sqrt{\hat X}}^2,
\ee
The variations are now easily obtained by defining $Y=\sqrt{\hat{X}}$, and writing
\be\label{GR'''}
S_{\rm GR}[A] =- \frac{2}{3} M^2_p M^2 \int \left( \mathrm{Tr} Y  \right)^2,
\ee
where we have dropped the integration measure $d^4x \sqrt{-g}$ for brevity. The variations are then easily computed:
\be
\delta S_{\rm GR}[A]=-\frac{2}{3} M^2_p M^2 \int  2   \  \Tr{Y}   \Tr{  \delta Y} , \\
 \delta^2 S_{\rm GR}= -\frac{2}{3} M^2_p M^2 \int  2 \ \left[  \ \Tr{ \delta Y} \Tr{ \delta Y}    +\Tr{Y}     \  \Tr{ \delta^2 Y }\right], \\
 \delta^3 S_{\rm GR}= -\frac{2}{3} M^2_p M^2 \int  2 \ \left[  3 \  \Tr{ \delta Y} \Tr{ \delta^2 Y} 
  +\Tr{Y}     \  \Tr{ \delta^3 Y}\right], \\
\delta^4 S_{\rm GR}=-\frac{2}{3} M^2_p M^2  \int  2 \ \left[  3  \Tr{\delta^2 Y}  \Tr{\delta^2 Y} +4 \Tr{\delta Y}  \Tr{\delta^3 Y}   +\Tr{Y}   \  \Tr{\delta^4 Y}\right].
\ee

It thus remains to obtain a relation between the variations of $Y$ and those of $\hat{X}$. This is easily done by varying the relation $Y^2=\hat{X}$ (any required number of times), and then solving the resulting equations for $\delta^k Y$. We only need these variations on the background, where we have $Y^{ij}\hat{=}\delta^{ij}$. This procedure gives:
\begin{align}
  & \delta Y\hat{=}  \frac{1}{2}\delta \hat{X},
\\
  & \delta^2 Y \hat{=} \frac{1}{2} \delta^2 \hat{X} - \delta Y\delta Y =\frac{1}{2}\left( \delta^2 \hat{X} -\frac{1}{2}\delta \hat{X} \delta \hat{X}  \right),
\\
  & \delta^3 Y  =  \frac{1}{2}\delta^3 \hat{X} -\frac{3}{2} \delta Y  \delta^2 Y- \frac{3}{2}  \delta^2 Y  \delta Y = \frac{1}{2}\delta^3 \hat{X} -\frac{3}{8}\left(   \delta^2 \hat{X}\delta \hat{X} +  \delta \hat{X}\delta^2 \hat{X}-\delta \hat{X}\delta \hat{X}\delta \hat{X}\right),
\\
  & \delta^4 Y  = -2 \delta Y  \delta^3 Y  - 2 \delta^3 Y \delta Y  - 6 \delta^2 Y  \delta^2 Y.
   \end{align}

The above results can be put into the general form (\ref{variations-action-general}) by writing:
 \be
 (3M^2/M_p^2) f^{(1)}_{ij}=3\delta_{ij},\\  \label{f2-GR}
 (3M^2/M_p^2)f^{(2)}_{ijkl}=-\frac{3}{2}P_{ijkl},\\ 
 (3M^2/M_p^2)f^{(3)}_{ijklmn} =\frac{9}{4} \sum_{\text perm} \frac{1}{3!} P_{ijab}P_{klbc}P_{mnca}+\frac{1}{2}\left(\delta_{ij} P_{klmn}+\delta_{kl} P_{ijmn}+\delta_{mn} P_{ijkl}\right)
 \\ \nonumber 
 (3M^2/M_p^2)f^{(4)}_{ijklmnpq}=-\frac{45}{8} \sum_{\text perm} \frac{1}{4!} P_{ijab}P_{klbc}P_{mncd}P_{pqda} + \frac{3}{8} \sum_{\text perm} \frac{1}{3} P_{ij kl} P_{mn pq} + \ldots,
 \ee
 where
 \be\label{P}
 P_{ijkl}:=\frac{1}{2}(\delta_{ik}\delta_{jl} +\delta_{il}\delta_{jk}) - \frac{1}{3}\delta_{ij}\delta_{kl}
 \ee
 is the projector on the symmetric tracefree matrices, and the dots in the last formula stand for terms containing at least one $\delta_{ij}$ in one of the 4 external "legs". The sum over permutations in the last two formulas is needed to make the result on the right-hand-side symmetric. Eventually we are going to contract $f^{(3)}, f^{(4)}$ with copies of the same matrix $\delta\hat{X}^{ij}$, and this sum over permutations (with the associated combinatorial factor) will disappear. Also, the reason why we don't write the remaining terms in the expression for $f^{(4)}$ is that (in the second paper from the series) we shall see that these terms will not play any role (in the 4-vertex) as they will be killed on-shell by the external states, or killed by the symmetries of the propagator when the vertices are used in Feynman graphs. 
 
 \subsection{Matrices $f^{(n)}_{ijkl...}$ for a general $f$}
 
For the case of a general theory we can to a large extent fix the derivatives of the function $f$ evaluated at the background $\hat{X}^{ij}=\delta^{ij}$ from the properties of $f$ itself. Thus, we know that $f$ is an ${\rm SO}(3)$ invariant function. The background that we work with is also ${\rm SO}(3)$ invariant. Thus, the same will be true for the matrices $f^{(n)}_{ijkl...}$. This, in particular, implies that the matrix of first derivatives must be proportional to $\delta_{ij}$. The proportionality coefficient can then be fixed from the homogeneity property of $f$ that implies
 \be\label{f-hom}
 \frac{\partial f}{\partial \hat{X}^{ij}} \hat{X}^{ij} = f.
 \ee
Thus, we have
\be
f^{(1)}_{ij} =  \frac{f(\delta)}{3} \delta_{ij}.
\ee
We also know from (\ref{action-backgr-general}) that $f(\delta)$ is the analog of the parameter $3M_p^2/M^2$ in GR for a general theory.

We can now differentiate the equation (\ref{f-hom}) once with respect to $\hat{X}^{ij}$ to obtain
\be
 \frac{\partial^2 f}{\partial \hat{X}^{ij}\partial \hat{X}^{kl}} \hat{X}^{ij} = 0.
 \ee
 In other words, the background itself is among the flat directions of the Hessian of $f$. This, together with the ${\rm SO}(3)$-invariance of the matrix $f^{(2)}_{ijkl}$ implies that it is of the form
 \be\label{f2}
 f^{(2)}_{ijkl} = -\frac{g}{2} P_{ijkl},
 \ee
where $g$ is some parameter and $P_{ijkl}$ is the projector (\ref{P}) introduced above. This must be true for any $f$. Note that this is also true for the function $f(\hat{X})\sim {\rm Tr}(\hat{X})$, i.e. for the topological theory, but in this case we have $g=0$. We shall see that there are propagating degrees of freedom whenever $g\not=0$. Finally, we note that we have put a minus sign in (\ref{f2}) because there is one in the case of GR, see (\ref{f2-GR}). It is natural to be interested in theories that are not too far from GR, and so it is natural to have the same sign in (\ref{f2}) as in GR. For this reason we shall assume $g>0$ in what follows. 
 
The higher derivatives $f^{(n)}_{ij\ldots}$ can all be determined in a similar fashion. Thus, one takes higher and higher derivatives of the equation (\ref{f-hom}) and evaluates the result on $\hat{X}^{ij}=\delta^{ij}$. One gets
\be\label{recurs}
f^{(n)}_{i_1 j_1 i_2 j_2 \ldots i_n j_n} \delta^{i_n j_n} + (n-2) f^{(n-1)}_{i_1 j_1 i_2 j_2 \ldots i_{n-1} j_{n-1}} =0,
\ee
which is a recursive relation for the matrices of derivatives. We see that the new independent term that appears at each order is always of the form of $n$ projectors (\ref{P}) contracted with each other in a loop, with a symmetrization over index pairs $ij$ later taken to form a completely symmetric expression. There are also terms where the projectors are contracted in smaller groups. Thus, we can write
\be
f^{(n)}_{i_1 j_1 i_2 j_2 \ldots i_n j_n} =  (-1)^{n-1} g^{(n)} \sum_{\text perm} \frac{1}{n!} P_{i_1 j_1 a_n a_1} P_{i_2 j_2 a_1 a_2} \ldots P_{i_n j_n a_{n-1} a_n} + \ldots,
\ee
where the dots denote terms that contain smaller groups of $P$ contractions, as we as terms that do not vanish when contracted with $\delta_{ij}$ in one of the channels. The coefficients in front of these latter terms are related to the lower $g^{(n)}$ via (\ref{recurs}). For example, for $f^{(3)}$ we have
\be
f^{(3)}_{ijklmn} =g^{(3)}\sum_{\text perm} \frac{1}{3!} P_{ijab}P_{klbc}P_{mnca}+\frac{g}{6}\left(\delta_{ij} P_{klmn}+\delta_{kl} P_{ijmn}+\delta_{mn} P_{ijkl}\right),
\ee
where $g\equiv g^{(2)}$. For the matrix of fourth derivatives we have
\be
 f^{(4)}_{ijklmnpq}=- g^{(4)} \sum_{\text perm} \frac{1}{4!} P_{ijab}P_{klbc}P_{mncd}P_{pqda} +\tilde{g}^{(4)} \sum_{\text perm} \frac{1}{3} P_{ij kl} P_{mn pq} \ldots,
 \ee
where the other terms contain at least one factor of $\delta$ and are not going to be important for us. Thus, the above parameterization of the derivatives of $f$ makes it clear that for a general theory there is an infinite number of independent coupling constants $g=g^{(2)}, g^{(3)}, \ldots$, with a number of new couplings appearing at each order of the derivative of the defining function. In turn, we could have chosen to parameterize $f$ by its independent couplings $g^{(n)}$. We (again) note that all these couplings are dimensionless.

We would like to emphasize that the procedure used to obtain the action variations is completely algorithmic and can be continued to arbitrary order without any difficulty.

\section{Free theory}
\label{sec:free}

The linearized action worked out below first appeared in \cite{Krasnov:2011up}, where also the Hamiltonian analysis (in the Minkowski limit) is contained. The novelty of this section is in the extension to the analysis to the more non-trivial de Sitter background. Also, the very compact form (\ref{S2-D}) of the completely symmetry reduced action is new. The most important new aspect of this section is in the realization that the connection cannot be taken to be real. This is invisible in the Minkowski version of the linearized action analysed in the previous works. Thus, our treatment of the reality conditions corrects and supersedes what appeared earlier in \cite{Krasnov:2011up} and \cite{Krasnov:2012pd}.

\subsection{Linearized Lagrangian}

In this paper we only consider the linearized theory. The second order action (obtained as $1/2$ of the second variation) reads: 
\be\nonumber
S^{(2)}=\int\left[\frac{g}{2}P_{ijkl} \Sigma^{i\mu\nu}D_\mu\delta A^{j}_\nu\Sigma^{k\rho \sigma}D_\rho\delta A^{l}_\sigma- \frac{f(\delta)}{3} \left(\frac{1}{\im}\epsilon^{\mu\nu\rho\sigma}D_\mu\delta A^i_{\nu}D_\rho\delta A^i_\sigma-M^2\Sigma^{i\mu\nu} \epsilon^{ijk}\delta A_\mu^j   \delta A_\nu^k\right)\right].
\ee
We first note that we can integrate by parts in the second term, with the result canceling the last term precisely. One uses (\ref{comm-DD}) to verify this. The integration by parts is justified on connection perturbations of compact support (in both space and time directions), and this is what we assume. Let us also absorb the prefactor $-g$ into the connection perturbation and define a new (canonically normalized as will be verified later) field
\be
a_\mu^i := \frac{\sqrt{g}}{\im} \delta A_\mu^i.
\ee
The free theory Lagrangian takes the following simple form:
\be\label{free}
{\cal L}^{(2)}=-\frac{1}{2}P_{ijkl} \Sigma^{i\mu\nu}D_\mu a^{j}_\nu \Sigma^{k\rho \sigma}D_\rho a^{l}_\sigma.
\ee
In this section we study this theory in some details. We start by listing the symmetries of the theory.

\subsection{Symmetries}

The free theory (\ref{free}) is invariant under the following local symmetries:
\be\label{lin-gauge}
\delta_\phi a_\mu^i = D_\mu \phi^i \qquad {\rm (gauge)}, \qquad 
\delta_\xi a_\mu^i = \xi^\alpha \Sigma_{\mu\alpha}^i \qquad {\rm (diffeo)}.
\ee
Note that the action of diffeomorphisms in this language is very simple, and corresponds to mere shifts of the connection in some directions. The first formula here is the usual action of the gauge symmetry. The second formula follows by writing the action of diffeomorphisms (modulo a gauge transformation) on $a_\mu^i$ as (\ref{intr-diffeos}), and then using the equation (\ref{F-S}) for the background curvature. The vector field appearing in (\ref{lin-gauge}) is then an appropriately rescaled one (by $M^2$) as compared to (\ref{intr-diffeos}). 

The invariance under the usual gauge rotations is easy to see using the result for the commutator of two covariant derivative (\ref{comm-DD}) and then the algebra (\ref{S-algebra}) of $\Sigma$-matrices. To verify the invariance under diffeomorphisms we use the fact that $D_{[\mu} \Sigma_{\nu\rho]}^i=0$ (this follows from (\ref{F-S}) and the Bianchi identity for the curvature). Writing this identity as
\be
D_{[\rho}\Sigma^i_{\sigma]\alpha} = -\frac{1}{2} D_\alpha \Sigma_{\rho\sigma}
\ee
the variation of the Lagrangian (\ref{free}) becomes:
\be\label{diff-1}
\delta_\xi {\cal L}^{(2)} = - P_{ijkl} \Sigma^{i\mu\nu}D_\mu a^{j}_\nu \left( -\frac{1}{2} \Sigma^{k\,\rho\sigma} \xi^\alpha D_\alpha \Sigma^l_{\rho\sigma}\right).
\ee
Here we have used the fact that in the term where the covariant derivative acts on the $\xi$ field and the $\Sigma$ matrix is taken outside of the sign of the derivative, the algebra of the $\Sigma$-matrices gives an expression that is either anti-symmetric in $\delta^{kl}$ or a pure trace. Both are killed by the projector $P_{ijkl}$, and so only the term present in brackets in (\ref{diff-1}) remains. But now we note that the expression in the brackets can be replaced with 
$$ -\frac{1}{4} \xi^\alpha D_\alpha \left( \Sigma^{k\,\rho\sigma} \Sigma^l_{\rho\sigma} \right)$$
in view of the $kl$-symmetrization implied by the projector. This expression, however, is proportional to the covariant derivative of the Kronecker $\delta$ in view of the algebra satisfied by $\Sigma$'s, and this is zero. This establishes the invariance under diffeomorphisms as well.

\subsection{Hamiltonian analysis}

We now follow the textbook procedure of the Hamiltonian analysis of (\ref{free}), to prepare the theory for the canonical quantization. Unlike what was done in \cite{Krasnov:2011up} we would like to remain in de Sitter background and not take the $M\to 0$ limit, at least not at this stage. We shall see that many subtleties, including those of the reality conditions, can only be understood for a non-zero value of $M$. So, we live in the de Sitter space (\ref{dS-metric}), with the self-dual two-forms given by (\ref{Sigmas}). We will also need a convenient expression for the background connection (\ref{A*}), and this is given by
\be\label{A'}
A_\mu^i = \frac{a'}{\im a} (dx^i)_\mu \equiv (\Hc/\im) (dx^i)_\mu,
\ee
where the prime denotes the (conformal) time derivative and we have introduced $\Hc=a'/a$. The equation (\ref{F-S}), which is just the Einstein equation(s) in our language, then states $\Hc'=\Hc^2=M^2 a^2$, with the solution being $a(t)= - 1/M(t-t_0)$, where $t_0$ is an arbitrary integration constant. 

We now compute the quantity $\Sigma^{i\,\mu\nu} D_\mu a_\nu^j$ in terms of the temporal $a_0^j$ and spatial $a_i^j$ components of the connection. We get:
\be\label{ham-1}
a^2 \Sigma^{i\,\mu\nu} D_\mu a_\nu^j = -\im \partial_t a^{ij}+ \im D^i a_0^j + \epsilon^{ikl}D_k a_l^j,
\ee
where $D_i$ is the covariant derivative with respect to the background connection (\ref{A'}). Explicitly
\be
D_k a_l^i = \partial_k a_l^i -\im \Hc \epsilon^{ikm} a_l^m,
\ee
where we have used (\ref{A'}). The convention in (\ref{ham-1}) is that the first index of $a^{ij}$ is the spatial one. 

We now decompose the spatial connection in its irreducible components
\be
a^{ij} = \tilde{a}^{ij} + \epsilon^{ijk} c^k + \delta^{ij} c,
\ee
where $\tilde{a}^{ij}$ is the symmetric tracefree component (i.e. spin 2). We substitute this into (\ref{ham-1}) and immediately find that the spin zero component $c$ gets projected away by the projector $P_{ijkl}$ that multiplies this quantity in the Lagrangian. Keeping only the symmetric tracefree parts we get
\be
a^2 P \Sigma^{i\,\mu\nu} D_\mu a_\nu^j = -\im \partial_t \tilde{a}^{ij} + \im \partial^i (a_0^j+\im c^j) + \epsilon^{ikl}\partial_k \tilde{a}_l^j + \im \Hc \tilde{a}^{ij}.
\ee
We see that the dependence on the anti-symmetric part $c^i$ can be absorbed into a shift of the temporal part. We therefore see that only the spin 2 part $\tilde{a}^{ij}$ of the spatial connection is dynamical. We drop the tilde from now on. The conjugate momentum to $a^{ij}$ is
\be
\pi^{ij}= \partial_t a^{ij}-P \partial^i (a_0^j+\im c^j) +\im B^{ij} -\Hc a^{ij},
\ee
where we have introduced the "magnetic" field $B^{ij}=P \epsilon^{(ikl}\partial_k a_l^{j)}$, where $P$ everywhere is the symmetric tracefree projector. The action in the Hamiltonian form becomes:
\be
S^{(2)}=\int dt \int d^3x \left( \pi_{ij} \partial_t a^{ij}- H \right),
\ee
where the Hamiltonian density is
\be
H=\frac{1}{2}\pi_{ij} \pi^{ij} -\im \pi_{ij} B^{ij} +\Hc \pi_{ij} a^{ij} - (a_0^i + \im c^i ) \partial^j \pi_{ij} .
\ee
We have integrated by parts in the Gauss constraint term. Note that all instances of the conformal factor $a$ have cancelled from the action. Indeed, we had a factor of $a^4$ coming from the measure $\sqrt{-g}$, as well as a factor of $a^{-2}$ twice coming from $\Sigma$'s with the raised spacetime indices. 

\subsection{Gauge-fixing}

It is convenient to fix the gauge at an early stage, and work with only the physical propagating modes. We see that the variation of the action with respect to the Lagrange multiplier $a_0^i$ gives the Gauss constraint
\be
\partial_i \pi^{ij} = 0.
\ee
This constraint generates gauge transformations
\be
\delta a^{ij} = P \partial_{(i} \xi_{j)},
\ee
where the projection is taken onto the tracefree part. This action can be used to set to zero the transverse part of $a^{ij}$:
\be
\partial_i a^{ij}=0,
\ee
which is our gauge-fixing condition. Thus, our dynamical fields are a pair $(a_{ij}, \pi^{ij})$ of symmetric traceless transverse tensors, as is appropriate for a spin 2 particle. We now note that the quantity $\epsilon^{ikl} \partial_k a_l^j$ is automatically symmetric tracefee and transverse on $a^{ij}$ that are symmetric tracefree and transverse. Thus, the projector in the definition of $B^{ij}$ can be dropped.

\subsection{Convenient notation}

The first-order differential operator $a^{ij}\to \epsilon^{ikl} \partial_k a_l^j $ acts on the space of symmetric tracefree transverse tensors. It will appear on many occasions below, and so it is convenient to introduce a special notation for it
\be
(\epsilon\partial a)^{ij} : = \epsilon^{ikl} \partial_k a_l^j .
\ee
It is then not hard to show that
\be\label{ed2}
(\epsilon\partial)^2 = -\Delta.
\ee
It is also not hard to see that $\epsilon\delta$ is self-adjoint with respect to the scalar product
\be
(x,y) = \int d^3 x \, x^{ij} y_{ij}
\ee
on the space of symmetric tracefree transverse tensors $x^{ij}, y^{ij}$. Then, using the self-adjointness and (\ref{ed2}) we can write the Hamiltonian as
\be\label{H-2}
H = \frac{1}{2} \pi^2 - \im\pi (\epsilon\partial a +\im \Hc a),
\ee
where we omitted the indices for brevity.

\subsection{Evolution equation}

Let us introduce two first order differential operators that are going to play an important role below. We define
\be\label{D-ops}
D := -\im \partial_t + \epsilon\partial + \im \Hc, \qquad
\bar{D} := \im \partial_t + \epsilon\partial + \im \Hc,
\ee
where $\bar{D}$ is clearly the adjoint of $D$ with respect to scalar product that also involves the time integration. We note that $Da$ is essentially the projected quantity $a^2 P \Sigma^{i\mu\nu} D_\mu a_\nu^j$, with the gauge-fixed spatial connection and its conjugated momentum satisfying the Gauss equation.

The Hamiltonian (\ref{H-2}) then results in the following Hamilton equations 
\be
-\im \pi = Da, \qquad \bar{D} \pi = 0,
\ee
which immediately give
\be\label{evol-a}
0= \bar{D} D a = \partial_t^2 a - \Delta a + 2\im \Hc \epsilon\partial a - 2\Hc^2 a
\ee
as the evolution equation. Because of the term with $\epsilon\partial$ that has a factor of $\im$ in front, this equation is complex. It becomes a non-trivial problem to choose a reality condition that is compatible with the evolution. Indeed, the naive reality condition that $a^{ij}$ is real is not consistent with the evolution, because if one starts with a real $a^{ij}$, the evolution will generate an imaginary part. Thus, a more sophisticated strategy for dealing with this problem is needed.

\subsection{Second-order formulation}

Let us rewrite the original action (\ref{free}) as a functional on the space of symmetric tracefree transverse tensors $a^{ij}$. This can also be obtained by integrating out the momentum variable. Using the operators (\ref{D-ops}) the corresponding second-order action can be written very compactly as
\be\label{S2-D}
S^{(2)} = - \frac{1}{2} \int dt\int d^3 x \,(Da)^2,
\ee
with (\ref{evol-a}) following immediately as the corresponding Euler-Lagrange equation.

\section{Reality conditions}
\label{sec:reality}

Our treatment of the connection field reality conditions in this section is new. This analysis constitutes one of the most important new results of this paper.

\subsection{Evolution equation as an eigenfunction equation}

For our later purposes, it is very convenient to write the evolution equation (\ref{evol-a}) in a slightly different form. Thus, we use the fact that
\be
\{ D,\bar{D}\} = 2\Hc^2,
\ee
which easily follows from $\Hc'=\Hc^2$, and write the evolution equation as an eigenfunction equation
\be\label{evol-D}
E a = a, \qquad {\rm where} \qquad E = \frac{1}{2\Hc^2} D\bar{D}.
\ee
This is the form that is going to be most useful below. 

\subsection{An important identity}

We now prove an identity that lies at the root of the reality condition that is going to be imposed. First, we note that
\be\label{ident-DH}
\bar{D} \frac{1}{2\Hc^2} = \frac{1}{2\Hc^2} D^*,
\ee
where $D^*= \im \partial_t + \epsilon\partial - \im \Hc$ is the operator complex conjugate to $D$. The above identity allows us to pull out a factor of $1/2\Hc^2$ from the derivative operator $\bar{D}$, and the expense of introducing a complex conjugate of $D$. 

We now consider the square of the evolution equation operator $E$:
\be
E^2 = \frac{1}{2\Hc^2} D\bar{D} \frac{1}{2\Hc^2} D\bar{D}.
\ee
We use (\ref{ident-DH}) to convert $\bar{D}$ into $D^*$ and then use the fact that $D$ and $D^*$ commute $\{ D,D^*\}=0$. We then use the complex conjugate of the identity (\ref{ident-DH}). Overall, we get the following sequence of transformations
\be\label{E2}
E^2 = \frac{1}{2\Hc^2} D \frac{1}{2\Hc^2} D^* D\bar{D} = \frac{1}{2\Hc^2} D \frac{1}{2\Hc^2} D D^* \bar{D} =\frac{1}{2\Hc^2} D \bar{D}^*\frac{1}{2\Hc^2}  D^*\bar{D} = R R^*,
\ee
where we have introduced
\be
R := \frac{1}{2\Hc^2} D \bar{D}^*.
\ee
Note that $R$ is a dimensionless operator, since $\Hc$ carries the dimension of mass. The identity (\ref{E2}) in particular implies that $E^2$ is a real operator, which is not at all obvious because $E$ is not real. 

\subsection{The reality condition}

In the case of the Dirac equation viewed as a reality condition for the spinors satisfying the Klein-Gordon equation, the Dirac equation appears as a "square root" of the Klein-Gordon. In our case we expect a second-order in derivatives reality condition, as follows from our general discussion in the Introduction. Thus, if it is to appear as a square root, it must be a square root of some fourth-order differential equation. 

Now, as our relation (\ref{E2}) demonstrtes, in spite of the fact that the evolution equation (\ref{evol-D}) is complex, we see that its square $E^2 a= a$, which is clearly implied by (\ref{evol-D}), is a real equation. This fourth order equation is not so interesting in itself, but introduces a new second-order differential operator $R$, such that $E^2=R R^*$. In other words, $R$ is a "square root" of the real equation operator $E^2$, similar to the Dirac operator being a square root of the Klein-Gordon one. It is then clear that if we define
\be
{\cal R}=R\circ \dagger,
\ee
which should be compared with (\ref{R-2}) in the Introduction, then the reality condition 
\be\label{reality-a}
{\cal R}a = a
\ee
is compatible with the evolution equation $Ea=a$. Indeed, the compatibility is just a rephrasal of the statement that on solutions of (\ref{evol-D}) the ${\cal R}$ anti-linear operator becomes an involution: 
\be\label{invol}
{\cal R}^2 = R R^* = E^2 ={\rm Id},
\ee
where the last equation holds on the space of solutions $Ea=a$. Thus, ${\cal R}$ is a real structure on the space of solutions, and the condition (\ref{reality-a}) is a possible reality condition that can be imposed. Below we shall see that this is the physically correct condition, in particular by working out a relation to the metric description. The essence of (\ref{reality-a}) will then be just a statement that the metric is real.

It is worth emphasizing that all of the above happens in exact analogy with the case of Dirac equation, except that now the relevant "Dirac" operator is second order, and appears as a square root of the fourth-order operator obtained by squaring the evolution operator. This squaring of the evolution equation procedure is absent in the fermionic case, where the condition that the square of the ${\cal R}$ operation is an identity is identical to the evolution equation. In our case this is not possible because the involution condition is necessarily fourth-order, and so it must be related to the evolution operator in a more non-trivial way (\ref{invol}).

\subsection{Metric}

We can now rephrase the condition (\ref{reality-a}) as a statement that a certain quantity is real. Indeed, we introduce 
\be\label{h-a}
h = \frac{1}{\sqrt{2}M} \bar{D} a,
\ee
where the prefactor is introduced for convenience and also in order to give $h$ the same mass dimension as $a$. Below we will show that $h$ can be viewed as just a possible new configuration variable on the phase space of the theory, with the Hamiltonian form action principle in terms of this variable taking an explicitly real form (\ref{H-h}). 

The evolution equation in its form (\ref{evol-D}) can now be rephrased by saying that it gives the inverse relation
\be\label{a-h}
a = \frac{M}{\sqrt{2}H^2} Dh.
\ee
Taking now the hermitian (complex) conjugate of the quantity $h$ in (\ref{h-a}), requiring it to be real
\be
h^\dagger = h,
\ee
and then substituting $h= \bar{D}a/\sqrt{2} M$ into (\ref{a-h}) we get precisely the reality condition (\ref{reality-a}). Thus, the essence of the condition (\ref{reality-a}) imposed on the space of solutions $Ea=a$ of our theory is indeed in the statement that the quantity (\ref{h-a}) is real. We note that this interpretation of the reality condition in terms of some quantity being real is not present in the case of the Dirac equation. Such an interpretation became possible because our reality condition is second order in derivatives, unlike the first order Dirac equation (=reality condition).

\subsection{Evolution equation for the metric}

As the last result of this section, let us use the identities derived above to obtain an evolution equation for the variable $h$. It is not hard to see that this equation is
\be\label{evol-h}
\frac{1}{2\Hc^2} D^* D h = h.
\ee
Indeed, using (\ref{ident-DH}) we can rewrite this as
\be
\bar{D} \frac{1}{2\Hc^2} D h = h \qquad {\rm or} \qquad \bar{D} \frac{1}{2\Hc^2} D \bar{D} a = \bar{D} a,
\ee
where to obtain the last equation we have used the relation (\ref{h-a}). The equation obtained is just the evolution equation $Ea=a$ with the operator $\bar{D}$ applied to it. Thus, (\ref{evol-h}) clearly follows from (\ref{evol-a}). It is also worth noting that it is a real equation, as is appropriate for a quantity that can consistently be assumed to be real. 

\section{Canonical transformation to the metric variables}
\label{sec:metric}

The purpose of this section it to explicitly carry out the field redefinition (\ref{h-a}) and see that it can get completed (once the momentum variable is considered) into a canonical transformation on the phase space of the theory. The content of this section is new. 

\subsection{Canonical transformation - momentum shift}

It is very convenient to eliminate the $\pi a$ cross-term in (\ref{H-2}) by shifting the momentum. Thus, we define
\be
\tilde{\pi}= \pi- \im (\epsilon\partial + \im \Hc) a.
\ee
Because of the last, time dependent (via $\Hc$) term the transformation of the symplectic form gives rise to a contribution to the Hamiltonian. In other words, modulo surface terms we get
\be
\pi \partial_t   = \tilde{\pi} \partial_t a + \frac{\Hc^2}{2} a^2,
\ee
where we have used $\Hc'=\Hc^2$. We now drop the tilde from the momentum variable, and write the reduced action in the Hamiltonian form as
\be
S^{(2)} = \int dt \int d^3 x \left( \pi\partial_t a - H\right),
\ee
with the Hamiltonian given by
\be\label{H-1}
H = \frac{1}{2} \pi^2 + \frac{1}{2} \left( \epsilon\partial a + \im \Hc a\right)^2 -  \frac{\Hc^2}{2} a^2.
\ee
The convenience of the new momentum variable lies in the fact that
\be
\partial_t a = \pi.
\ee

\subsection{Canonical transformation to $h$ variables}

From the previous section we know that we should be able to describe the dynamics in terms of the variable
\be
h = \frac{1}{\sqrt{2} M} \left( \im \pi + (\epsilon\partial + \im \Hc) a\right),
\ee
and that this variable can consistently be assumed to be real. The canonically conjugate momentum $p$ to $h$ is of course only defined modulo $a$-dependent shifts. However, if we insist that there is no $p h$ terms in the resulting Hamiltonian, then the momentum variable can be determined to be given by
\be
p = \frac{M}{\sqrt{2}\Hc^2} \left( (\epsilon\partial + \im \Hc) \pi - \im \left( (\epsilon\partial + \im \Hc)^2 - 2\Hc^2\right) a\right.
\ee
We emphasize that this is a linear canonical transformation on the phase space of the theory.

\subsection{Metric Hamiltonian}

There are many contributions from the symplectic $\pi \partial_t a$ term to the Hamiltonian in terms of $h,p$ variables. After a rather tedious computation one finds that the action can be written as
\be
S^{(2)} = \int dt \int d^3x \left( p\partial_t h - H\right),
\ee
where
\be\label{H-h}
H = \frac{\Hc^2}{2M^2} p^2 + \frac{(\epsilon\partial)^2 - 2\Hc^2}{2\Hc^2} M^2 h^2. 
\ee
As a check, we note that this Hamiltonian goes into that for a massless field in the limit $M\to 0$. Indeed, using the explicit expression (\ref{Hubble}) for $\Hc$ one sees that $\Hc/M\to 1$ when $M\to 0$. This shows that the above Hamiltonian has the correct Minkowski limit. As for the de Sitter Hamiltonian, the above is the standard Hamiltonian for the de Sitter space spin 2 part of the metric perturbation $h_{\mu\nu}$ rescaled by the conformal factor $c(t)$. 

\subsection{Second-order formulation}

It is also instructive to write the above action in the second-order form, by integrating $p$ out. We get
\be\label{S-h}
S^{(2)} = - M^2 \int dt \int d^3x \,\frac{h}{2\Hc^2} \left( D^* D  - 2\Hc^2 \right) h = - M^2 \int dt \int d^3x \left( \frac{1}{2\Hc^2}  (Dh)^2 - h^2 \right),
\ee
where we have integrated by parts in the $(\partial_t h)^2$ term to get the first expression for the action, which is explicitly real, and have used (\ref{ident-DH}) to get the second, more symmetric expression. The first version of the action clearly leads to (\ref{evol-h}) as the corresponding Euler-Lagrange equation.

It is worth emphasizing that the connection formalism linearized action (\ref{S2-D}) is actually simpler than the same action (\ref{S-h}) in the metric description. Here we are comparing only the completely symmetry reduced actions, but the same holds true also about the full linearized Lagrangians in the two formulations. The graviton gauge-theoretic Lagrangian (\ref{free}) is much simpler than its metric variant. And, although we do not discuss it in any length in this paper, the connection Lagrangian (\ref{free}) (in its Euclidean signature version where all fields are real) is actually a non-negative function in the space of fields, which is not the case for the Euclidean signature metric Lagrangian because of the conformal mode. We will give a more detailed comparison of the off-shell Lagrangians in the second paper of the series, when we work out the propagator. 

\section{Canonical quantization and the mode decomposition}
\label{sec:modes}

We now perform all the usual steps for the canonical quantization of the theory (\ref{S2-D}), with the reality condition (\ref{reality-a}). Our main aim is to obtain a mode decomposition with correctly normalized creation and annihilation operators. The content of this section is new. 

\subsection{Choice of the time coordinate}

We first explicitly solve the evolution equation (\ref{evol-a}) for the connection, so that the linearly independent solutions later become the modes of the field. For this, let us first introduce a convenient parameterization of the $c(t)$ and $\Hc$ functions. We choose
\be
c(t) = \frac{1}{1-Mt}
\ee
so that $c(0)=1$, i.e. we have chosen to origin of the time coordinate in such a way that $t=0$ corresponds to the conformal factor of unity. With this parameterization we get
\be\label{Hubble}
\Hc = \frac{M}{1-Mt}.
\ee

\subsection{Spatial Fourier transform}

We now perform the spatial Fourier transform, and choose convenient polarization tensors. Thus, consider a mode of the form $a_k^{ij} e^{\im\vec{k}\vec{x}}$. The transverse condition $\partial_i a^{ij}$ on the connection implies that the corresponding mode $a_k^{ij}$ is orthogonal to $k^i$. For this reason, it is very convenient to define
\be
z^i(k) := k^i/|k|,
\ee
i.e. a unit vector in the direction of the spatial momentum. We then define two (complex) vectors $m^i(k), \bar{m}^i(k)$ that are both orthogonal to $z^i$ and whose only non-zero scalar product is $m^i \bar{m}_i = 1$. They satisfy 
\be
\im \epsilon^{ijk} z_j m_k=m_i, \qquad \im \epsilon^{ijk} z_j \bar{m}_k=-\bar{m}_i, \qquad \im \epsilon^{ijk}m_j \bar{m}_k=z_i.
\ee
Here we have omitted the momentum dependence of these vectors for brevity, but it should all the time be kept in mind that they are $\vec{k}$ dependent. Thus, when we replace $\vec{k}\to -\vec{k}$ the vectors $m^i, \bar{m}^i$ get interchanged:
\be
m^i(-k) = \bar{m}^i(k), \qquad \bar{m}^i(-k) = m^i(k).
\ee
It is very important to keep these transformations in mind for the manipulations that follow. 

\subsection{Polarization tensors}

The fact that $a^{ij}$ is symmetric tracefree transverse implies that every mode $e^{\im\vec{k}\vec{x}}$ comes in just two polarizations. For the corresponding polarization tensors it is convenient to choose $m^i(k) m^j(k)$ and $\bar{m}^i(k) \bar{m}^j(k)$. We shall refer to the $mm$ mode as the {\it negative} helicity particle, while the $\bar{m}\bar{m}$ mode will be referred to as the positive one. We will explain a reason for this choice below. 

Let us now consider the action of the operator $\epsilon\partial$ on the two polarizations. We have 
\be
(\epsilon\partial) m^i m^j a_k^- e^{\im\vec{k}\vec{x}} = \omega_k m^i m^j a_k^- e^{\im\vec{k}\vec{x}}, \qquad
(\epsilon\partial) \bar{m}^i \bar{m}^j a_k^+ e^{\im\vec{k}\vec{x}} = - \omega_k \bar{m}^i \bar{m}^j a_k^+ e^{\im\vec{k}\vec{x}},
\ee
where we have introduced
\be
\omega_k := |k|.
\ee
In other words, the two modes we have introduced are the eigenvectors of the operator $\epsilon\partial$ with eigenvalues $\pm\omega_k$ respectively. Our choice of the name for the $mm$ mode as negative may seem unnatural at the moment (since it is the positive sign eigenvalue of $\epsilon\partial$). However, it becomes more natural if one computes the corresponding Weyl curvatures for the two modes. One finds that the negative mode has zero self-dual Weyl curvature, and is thus a purely anti-self-dual object. This is why it makes sense to refer to it as the negative helicity mode.  

\subsection{Linearly independent solutions}

We now write the evolution equation (\ref{evol-a}) as an equation for the time evolution of the Fourier coefficients. We get, for each of the modes
\be
\partial_t^2 a_k^- + (\omega_k^2  + 2\im \Hc \omega_k - 2\Hc^2) a_k^- =0, \qquad
\partial_t^2 a_k^+ + (\omega_k^2  - 2\im \Hc \omega_k - 2\Hc^2) a_k^+ =0.
\ee
Note that the positive helicity equation is just the complex conjugate of the negative helicity one. 

Each of the above equations is a second order ODE, and thus has a positive and negative frequency solutions. It is not at all hard to obtain then explicitly, and they read
\be\label{modes}
a_k^- \sim \Hc e^{- \im\omega_k t} , \qquad a_k^- \sim \frac{1}{\Hc} e^{\im \omega_k t} \left( 1 - \frac{\im \Hc}{\omega_k} - \frac{\Hc^2}{2\omega_k^2} \right), \\ \nonumber
a_k^+ \sim \frac{1}{\Hc} e^{- \im\omega_k t} \left( 1 + \frac{\im \Hc}{\omega_k} - \frac{\Hc^2}{2\omega_k^2} \right), \qquad a_k^+ \sim \Hc e^{\im \omega_k t}.
\ee
It is interesting to note that one of the modes in each case is given by a rather simple expression, with the time-dependence of the amplitude being just that of $\Hc$. The other mode in each case is more involved. For the negative mode it is the positive frequency solution that is simple, while for the positive mode the positive frequency solution is involved. This is a manifestation of a general pattern in our formalism, in that the negative helicity mode will always be much easier to deal with then the positive helicity one. 

Another point worth emphasizing is that one of the two linearly independent solutions of the connection evolution equation is actually simpler than the modes in the metric description, see (\ref{h-mode-dec}) below. This gives yet another illustration of the general statement that we would like to promote - the connection description is in many aspects simpler than the metric one.

\subsection{Action of the $\bar{D}$ operator on the modes}

It is useful to compute the action of the basic operator $\bar{D}$ on the modes (\ref{modes}). We will need this when we impose the reality condition (\ref{reality-a}), which can be written as $a=(1/2\Hc^2) D (\bar{D} a)^\dagger$. We have
\be\label{bD-a}
\bar{D} m^i m^j \Hc e^{- \im\omega_k t+\im\vec{k}\vec{x}}= 2\omega_k m^i m^j \Hc e^{- \im\omega_k t+\im\vec{k}\vec{x}} \left(1+ \frac{\im \Hc}{\omega_k} \right), \\ \nonumber
\bar{D} \bar{m}^i \bar{m}^j \frac{1}{\Hc} e^{- \im\omega_k t+\im\vec{k}\vec{x}} \left( 1 + \frac{\im \Hc}{\omega_k} - \frac{\Hc^2}{2\omega_k^2} \right) = - \bar{m}^i \bar{m}^j \frac{\Hc}{\omega_k}  e^{- \im\omega_k t+\im\vec{k}\vec{x}} \left(1+ \frac{\im \Hc}{\omega_k} \right), \\ \nonumber
\bar{D} \bar{m}^i \bar{m}^j \frac{1}{\Hc} e^{\im\omega_k t-\im\vec{k}\vec{x}} \left( 1 - \frac{\im \Hc}{\omega_k} - \frac{\Hc^2}{2\omega_k^2} \right) = \bar{m}^i \bar{m}^j \frac{\Hc}{\omega_k} e^{\im\omega_k t-\im\vec{k}\vec{x}}\left(1- \frac{\im \Hc}{\omega_k} \right), \\ \nonumber
\bar{D} m^i m^j \Hc e^{\im\omega_k t-\im\vec{k}\vec{x}} = - 2\omega_k m^i m^j \Hc e^{\im\omega_k t-\im\vec{k}\vec{x}} \left(1- \frac{\im \Hc}{\omega_k} \right).
\ee
Now, to impose the reality condition, we take the complex conjugates of the right-hand-sides, and then apply the operator $D$ to them. We get
\be
2\omega_k D \bar{m}^i \bar{m}^j \Hc  e^{\im\omega_k t-\im\vec{k}\vec{x}} \left(1- \frac{\im \Hc}{\omega_k} \right) = (2\omega_k)^2 \bar{m}^i \bar{m}^j \Hc e^{\im\omega_k t-\im\vec{k}\vec{x}} \left( 1 - \frac{\im \Hc}{\omega_k} - \frac{\Hc^2}{2\omega_k^2} \right), \\ \nonumber
- D m^i m^j \frac{\Hc}{\omega_k}  e^{\im\omega_k t-\im\vec{k}\vec{x}} \left(1- \frac{\im \Hc}{\omega_k} \right)=m^i m^j \frac{\Hc^3}{\omega_k^2}  e^{\im\omega_k t-\im\vec{k}\vec{x}}, \\ \nonumber
D m^i m^j \frac{\Hc}{\omega_k} e^{-\im\omega_k t+\im\vec{k}\vec{x}} \left( 1 + \frac{\im \Hc}{\omega_k}\right) = m^i m^j \frac{\Hc^3}{\omega_k^2}  e^{-\im\omega_k t+\im\vec{k}\vec{x}}, \\ \nonumber
- 2\omega_k D \bar{m}^i \bar{m}^j \Hc e^{-\im\omega_k t+\im\vec{k}\vec{x}} \left(1+ \frac{\im \Hc}{\omega_k} \right)=(2\omega_k)^2 \bar{m}^i \bar{m}^j \Hc e^{-\im\omega_k t+\im\vec{k}\vec{x}} \left( 1 + \frac{\im \Hc}{\omega_k} - \frac{\Hc^2}{2\omega_k^2} \right).
\ee

\subsection{The mode expansion}

Using the above results, we can now write down a mode expansion satisfying the reality condition (\ref{reality-a}). We get
\be\label{a-mode-dec}
a^{ij}(t,\vec{x}) = \int \frac{d^3k}{(2\pi)^3 2\omega_k} \Big[ m^i m^j a_k^- \frac{\Hc}{\sqrt{2}\omega_k} e^{-\im\omega_k t+\im\vec{k}\vec{x}} + \bar{m}^i\bar{m}^j (a_k^-)^\dagger \frac{\sqrt{2}\omega_k}{\Hc} e^{\im\omega_k t-\im\vec{k}\vec{x}} \left( 1 - \frac{\im \Hc}{\omega_k} - \frac{\Hc^2}{2\omega_k^2} \right) \\ \nonumber
-\bar{m}^i\bar{m}^j a_k^+ \frac{\sqrt{2}\omega_k}{\Hc} e^{-\im\omega_k t+\im\vec{k}\vec{x}} \left( 1 + \frac{\im \Hc}{\omega_k} - \frac{\Hc^2}{2\omega_k^2} \right) - m^i m^j (a_k^+)^\dagger \frac{\Hc}{\sqrt{2}\omega_k} e^{\im\omega_k t-\im\vec{k}\vec{x}} \Big].
\ee
Here all the vectors $m^i,\bar{m}^i$ are $\vec{k}$-dependent, but this dependence is suppressed in order to have a compact expression. We could have chosen to put a plus sign in front of the positive helicity modes, but below we shall see that the above choice leads to a more symmetric expression for the metric mode expansion.

Note that the reality condition makes it unnatural to put factors of $M$ in front of the modes. Thus, as it stands, the expression (\ref{a-mode-dec}) does not have the Minkowski limit $M\to 0$, because some terms go to zero in this limit, and some other terms blow up. This is one difference with e.g. the Majorana fermion, which has a very similar type of the mode expansion. However, in that case there is a massless $m\to 0$ limit in which half of the modes are set to zero, but the other half survives and gives the mode expansion of the Weyl fermion. In our case the connection (\ref{a-mode-dec}) does not admit the $M\to 0$ limit. 

We also note that in (\ref{a-mode-dec}) only the relative coefficient between the $a, a^\dagger$ terms in each helicity sector is fixed by the reality condition, so we could have multiplied each sector by an arbitrary constant factor. By doing this we could obtain an expression that survives in the $M\to 0$ limit. However, we are now going to show that the mode decomposition (\ref{a-mode-dec}) is written in terms of canonically normalized operators. We do this by computing the commutators as implied by the canonical Poisson brackets between the connection and its conjugate momentum. 

\subsection{Commutators}

We start with the relation that the equal time connection and its conjugate momentum should satisfy:
\be\label{comm-ap}
[ a_{ij}(t,\vec{x}), \partial_t a_{kl}(t,\vec{y})] = \im \delta^{3}(x-y) P_{ijkl}.
\ee
For the conjugate momentum we have
\be
\partial_t a_{ij}(t,\vec{y}) = \int \frac{d^3p}{(2\pi)^3 2\omega_p} (-\im\omega_p) \Big[ m^i(p) m^j(p) a_p^- \frac{\Hc}{\sqrt{2}\omega_p} e^{-\im\omega_p t+\im\vec{p}\vec{y}} \left( 1+ \frac{\im\Hc}{\omega_p}\right) 
\\ \nonumber
- \bar{m}^i(p)\bar{m}^j(p) (a_p^-)^\dagger \frac{\sqrt{2}\omega_p}{\Hc} e^{\im\omega_p t-\im\vec{p}\vec{y}} \left( 1 - \frac{\Hc^2}{2\omega_k^2} +\frac{\im\Hc^3}{2\omega_k^3} \right) 
\\ \nonumber
-\bar{m}^i(p)\bar{m}^j(p) a_p^+ \frac{\sqrt{2}\omega_p}{\Hc} e^{-\im\omega_p t+\im\vec{p}\vec{y}} \left( 1 - \frac{\Hc^2}{2\omega_k^2} -\frac{\im\Hc^3}{2\omega_k^3} \right) 
\\ \nonumber
 + m^i(p) m^j(p) (a_p^+)^\dagger \frac{\Hc}{\sqrt{2}\omega_p} e^{\im\omega_p t-\im\vec{p}\vec{y}} \left( 1- \frac{\im\Hc}{\omega_p}\right) \Big].
\ee
Substituting this into (\ref{comm-ap}), and using the fact that under $\vec{k}\to -\vec{k}$ the vectors $m^i,\bar{m}^i$ get interchanged, as well as the fact that for any $\vec{k}$
\be
P_{ijkl} = m_i m_j \bar{m}_k \bar{m}_l + \bar{m}_i \bar{m}_j m_k m_l,
\ee
we get
\be
[a_k^\pm, (a_k^\pm)^\dagger]= (2\pi)^3 2\omega_k \delta^3(k-p),
\ee
which are the canonical commutational relations for the creation-annihilation operators in field theory. This gives one confirmation of the correct normalization used in (\ref{a-mode-dec}). Another confirmation comes by computing the metric, and then the associated Hamiltonian.

\subsection{Metric}

Let us now use (\ref{a-mode-dec}) to obtain the mode decomposition for the metric (\ref{h-a}). The action of the operator $\bar{D}$ on all the modes has already been computed in (\ref{bD-a}). We get
\be\label{h-mode-dec}
h^{ij}(t,\vec{x}) =  \frac{\Hc}{M} \int \frac{d^3k}{(2\pi)^3 2\omega_k} \Big[ (m^i m^j a_k^-  + \bar{m}^i\bar{m}^j a_k^+) e^{-\im\omega_k t+\im\vec{k}\vec{x}} \left( 1 + \frac{\im \Hc}{\omega_k}\right) \\ \nonumber
+ (\bar{m}^i\bar{m}^j (a_k^-)^\dagger + m^i m^j (a_k^+)^\dagger  ) e^{\im\omega_k t-\im\vec{k}\vec{x}} \left( 1 - \frac{\im \Hc}{\omega_k} \right) \Big].
\ee
This expression has an obvious (correct) Minkowski limit $M\to 0$. It is also explicitly hermitian. It is in order to obtain the above symmetric expression that we chose to introduce the minus signs in front of the positive helicity modes in (\ref{a-mode-dec}). To compute the Hamiltonian in terms of the modes, let us also give an expression for the momentum $p = (M^2/\Hc^2) \partial_t h$. We get
\be
p^{ij}(t,\vec{x}) =  \frac{M}{\Hc} \int \frac{d^3k}{(2\pi)^3 2\omega_k} (-\im \omega_k) \Big[ (m^i m^j a_k^- + \bar{m}^i\bar{m}^j a_k^+) e^{-\im\omega_k t+\im\vec{k}\vec{x}} \left( 1 + \frac{2\im \Hc}{\omega_k}-\frac{2\Hc^2}{\omega_k^2} \right) \\ \nonumber
- (\bar{m}^i\bar{m}^j (a_k^-)^\dagger+m^i m^j (a_k^+)^\dagger ) e^{\im\omega_k t-\im\vec{k}\vec{x}} \left( 1 - \frac{2\im \Hc}{\omega_k}-\frac{2\Hc^2}{\omega_k^2} \right) \Big].
\ee
The Hamiltonian (\ref{H-h}) then reads:
\be
\int H = \frac{1}{2} \int \frac{d^3k}{(2\pi)^3 2\omega_k} \omega_k \left( a_k^- (a_k^-)^\dagger + (a_k^-)^\dagger a_k^- + a_k^+ (a_k^+)^\dagger + (a_k^+)^\dagger a_k^+\right)\left( 1- \frac{\Hc^2}{2\omega_k^2} + \frac{\Hc^4}{\omega_k^4}\right).
\ee
The Hamiltonian is explicitly time dependent, as is appropriate for particles in time-dependent de Sitter Universe where the energy is not conserved. We note that it has the correct Minkowski limit $M\to 0$. 

\section{Discrete symmetries}
\label{sec:discrete}

In this section we obtain the action of the discrete C, P, T symmetries on the connection field, and on the creation-annihilation operators. 

\subsection{Charge conjugation}

Our fields are "real", in the sense that we do not have independent operators in front of the positive and negative frequency modes. The metric is explicitly real. Thus, the charge conjugation acts trivially - all operators go into themselves.

\subsection{Parity}

We could obtain the action of parity from the mode expansion for the metric, which is standard. We could also just directly define the action on the operators. Indeed, parity changes the sign of the spatial momentum, and interchanges the two helicities:
\be\label{parity}
P^\dagger a^\pm_k P = a_{-k}^\mp.
\ee
In view of (\ref{h-mode-dec}) this is equivalent to
\be
P^\dagger h^{ij}(t,\vec{x}) P = h^{ij}(t,-\vec{x}).
\ee
It is much more interesting to obtain the parity action on the connection field. Using (\ref{parity}) and the mode decomposition (\ref{a-mode-dec}) we get
\be\label{parity-a}
P^\dagger a^{ij}(t,\vec{x}) P = - (a^{ij}(t,-\vec{x}))^\dagger.
\ee
The minus sign in this formula can be interpreted as being related to the fact that we are dealing with the spatial connection, which changes sign under parity. But most importantly, we see that parity is related to the hermitian conjugation of the connection field operator. This is reminiscent of what happens in the case of fermions, where the parity at the level of 2-component spinors is also related to the hermitian conjugation of the spinor fields. 

\subsection{Time reversal}

Time-dependent physics in de Sitter space is not time reversal invariant. However, it can be made to be such by simultaneously reversing the sign of the time coordinate and the sign of the parameter $M$. This sends one from one patch of de Sitter space (covered by the flat slicing) to another patch where the time flows in the opposite direction. Hence, it must be a symmetry of the theory. The action of the time reversal, which is an anti-linear operator, can then be obtained by requiring
\be
T^\dagger h^{ij}(t,\vec{x}) T = h^{ij}(-t,\vec{x})\Big|_{M\to-M}.
\ee
This gives, at the level of the operators
\be\label{time-inv}
T^\dagger a^\pm_k T = a_{-k}^\pm.
\ee
While parity flips the sign of the spatial momentum while leaving the particle's spin unchanged, which results in flipping of the helicity, time reversal flips both the momentum and the spin, which does not change helicity. At the level of the connection we get
\be\label{time-a}
T^\dagger a^{ij}(t,\vec{x}) T = a^{ij}(-t,\vec{x})\Big|_{M\to-M}.
\ee

\subsection{CPT}

We now combine all of the above transformation rules into the action of the $CPT$ transformation. We see that, modulo an overall minus sign, this action is that of the spacetime inversion $(t,\vec{x})\to - (t,\vec{x})$, as well as the hermitian conjugation of the field. This is of course standard in field theory. Note, however, that in our case the hermitian conjugation comes not from the charge conjugation, in spite of the fact that the field is complex. Rather, it is a part of the parity transformation. But the end result is the same: $CPT$ is hermitian conjugation together with the spacetime inversion. This is the $CPT$ theorem for our theories - a {\it hermitian} Lagrangian will be $CPT$ invariant. At the same time, hermiticity of the Lagrangian is important for unitarity of the theory. While we have seen this hermiticity at the linearized level (e.g. by going to the metric description), the question whether there exists an appropriate real structure on the space of solutions of the full theory that allows a real section to be taken is open. 

\section{Discussion}

Let us recap the main points of our construction. We have studied diffeomorphism-invariant gauge theories of the type (\ref{action}) with the gauge group ${\rm SL}(2,\C)$, with the aim of describing the linearized theory around a background connection that corresponds to the de Sitter space. We have seen that all theories of this type coincide at the linearized level, and describe massless spin 2 particles. We have also seen that the arising connection evolution equation is in general complex, with the imaginary part appearing with a factor of the Hubble parameter $\Hc$ in front. Thus, in a time-dependent background such as the one given by the de Sitter space, the connection cannot be taken to be real. We also gave general arguments to the same effect based on the fact that the (linearized) connection realizes an intrinsically complex spinor representation $S_+^3\otimes S_-$ of the Lorentz group. At the same time, we have seen that a real structure exists on the space of solutions, and that this can be used to select a real section in the phase space, on which one obtains a theory with a hermitian Hamiltonian. All this was shown to be quite analogous to the treatment of fermions in which they are described as complex fields satisfying the Klein-Gordon equation, with an additional first order in derivatives reality condition (Dirac equation) imposed. The main difference with the case of fermions was that in our case the reality condition was necessarily of the second order in derivatives. We have also seen that this second order nature of the reality conditions is what guarantees that a real (metric) description exists. 

We have avoided discussing the above statements in the spacetime form, staying all the time at the level of the phase space formulation. On one hand this makes things more clear. On the other hand, for path integral computations it is necessary to develop the spacetime version of the mode decomposition. This will be accomplished in the second paper of the series, where this formalism is used to compute the graviton scattering amplitudes. One of the reasons why this was not treated already in the present paper is that it requires a much more detailed introduction into the spinor techniques (e.g. spinor helicity), and this would take us too far from the present goal of expanding the connection into the canonically normalized creation-annihilation operators. 

Let us finish with a very brief list of the open problems of this approach. The one that is most directly related to the topics covered in this paper is that of unitarity. Thus, it is not clear if there exists a satisfactory way to select a real section of the non-linear dynamics described by a general theory from the class (\ref{action}). However, the fact that this is possible in the linearized theory around such a time-dependent background as de Sitter, and the fact that at least for one of the theories from this class, namely GR, this is possible also at the full non-linear level, allows for optimism. 

The other major open problem of this approach is coupling to matter. Many types of bosonic matter can be coupled just by enlarging the gauge group, i.e. considering still theories of the same general class (\ref{action}), but with a larger $G\supset {\rm SL}(2,\C)$. In particular, Yang-Mills fields, as well as e.g. a massive scalar field can be coupled this way naturally. A very interesting symmetry breaking mechanism selecting what should be called the gravitational ${\rm SL}(2,\C)$ then becomes available, see \cite{Krasnov:2011hi} for more details. However, the arising matter/gravity dynamics should be studied in more details, in particular with the reality conditions issues in mind. An open question is that of coupling of fermions. This seems difficult in the usual first-order in derivatives formalism, but it should also be kept in mind that the fermions can also be described via a second-order in derivatives action, with a first-order reality condition imposed, as described in more details in the Introduction. This brings fermions much close to what seems to be at work in the class of theories considered here, and raises hopes that they can be coupled satisfactorily. 

The third major open problem of this approach is renormalizability. It has been conjectured in \cite{Krasnov:2006du} that the class (\ref{action}) with $G={\rm SL}(2,\C)$ is closed under renormalization. Work is in progress on testing this conjecture at one loop. Even if this turns out not to be the case for $G={\rm SL}(2,\C)$, it will still be possible that only for some specific choices of $G$ the class of theories (\ref{action}) becomes renormalization closed. For example, this may be the case when $G$ is an appropriate graded Lie group (i.e. a Lie supergroup). Such more general choices of $G$ may in any case be necessary to describe fermionic particles with their anti-commuting Grassmann-valued fields. These various version of the conjecture \cite{Krasnov:2006du} should be tested, and the formalism developed here for $G={\rm SL}(2,\C)$ is a necessary prerequisite for computations of this type. 

\section*{Acknowledgements} KK was supported by an ERC Starting Grant 277570-DIGT, as well as partially by a fellowship from the Alexander von Humboldt foundation, Germany. 

\appendix
\section{Appendix: Self-dual two-forms}

For any self-dual two-form we have:
\be
\frac{1}{2} \epsilon_{\mu\nu}{}^{\rho\sigma} U_{\rho\sigma} = \im U_{\mu\nu},
\ee
and for anti-self-dual form we have an extra minus on the right-hand-side. The space of self-dual two-forms being 3-dimensional, we can introduce a basis in it. A choice of such basic self-dual two-forms can be rather arbitrary as long as they span the required subspace. However, there is always a canonical (modulo certain gauge rotations, see below) choice of the basis. Let us denote such canonical basis self-dual two-forms by $\Sigma^i_{\mu\nu}, i=1,2,3$. Note that we have denoted the index enumerating the two-forms by the same letter as was used to refer to the spatial index in the Hamiltonian analysis. This is not an oversight; the two indices can be naturally identified, see below. The canonical basic self-dual two-forms are defined to satisfy 
\be\label{metricity}
\epsilon^{\mu\nu\rho\sigma} \Sigma^i_{\mu\nu}  \Sigma^j_{\rho\sigma}  = 8\im  \delta^{ij},
\ee
where the numerical coefficient on the right is convention-dependent, and $\delta^{ij}$ is the Kronecker-delta. It can be shown that the self-dual two-forms satisfying (\ref{metricity}) are defined uniquely modulo ${\rm SO}(3)$ rotations preserving $\delta^{ij}$. We can now give an explicit form of the basic self-dual two-forms in the case of the Minkowski spacetime metric. Using the two-form notation we have:
\be\label{Sigma}
\Sigma^i=\im dt\wedge dx^i + \frac{1}{2} \epsilon^{ijk} dx^j\wedge dx^k.
\ee
it is not hard to check the $\Sigma^i_{\mu\nu}$ are self-dual (with the conventions that $\epsilon^{0123}=+1$), and that (\ref{metricity}) holds. Let us also note what becomes of the components of the basis self-dual two-forms $\Sigma^i_{\mu\nu}$ under the space+time split. We have:
\be
\Sigma^i_{0j} = \im \, \delta^i_j, \qquad \Sigma^i_{jk} = \epsilon^{i}{}_{jk}.
\ee
Thus, we see that the objects $\Sigma^i_{\mu\nu}$ indeed provide a natural identification of the basis index $i$ with the spatial index. Let us also note an important identity satisfied by our self-dual two-forms. We have
\be\label{S-algebra}
\Sigma^i_\mu{}^\nu \Sigma^j_\nu{}^\rho = -\delta^{ij} \eta_\mu{}^\rho +\epsilon^{ijk} \Sigma^k_\mu{}^\rho.
\ee
Thus, the basic self-dual two-forms satisfy an algebra similar to that of Pauli matrices. This identity can be checked by direct verification, using the explicit expression (\ref{Sigma}).

\end{document}